\documentclass[11pt]{article}
\usepackage{epsfig,cite}
\usepackage{graphicx}
\usepackage{multicol}
\usepackage{color}
\usepackage{amssymb}

\def\bea{\begin{eqnarray}}
\def\eea{\end{eqnarray}}
\def\gluino{{\widetilde{g}}}
\newcommand\prd[3]   
		{{Phys.\ Rev.\ }{\bf D #1} (#2) #3}
\newcommand\prl[3]   
		{{Phys.\ Rev.\ Lett.\ }{\bf #1} (#2) #3}
\newcommand\plb[3]   
		{{Phys.\ Lett.\ }{\bf B #1} (#2) #3}
\newcommand\npb[3]    
		{{Nucl.\ Phys.\ }{\bf B #1} (#2) #3}
\newcommand\app[3]   
		{{Astropart.\ Phys.\ }{\bf #1} (#2) #3}
\newcommand\jhep[3]  
		{{J. High Energy Phys.\ }{\bf #1} (#2) #3}
\newcommand\epjc[3]  
		{{Eur.\ Phys.\ J. }{\bf C #1} (#2) #3}
\newcommand\npps[3]  
		{{Nucl.\ Phys.\ }{\bf #1} {\it(Proc.\ Suppl.)} (#2) #3}
\newcommand\jcap[3]  
		{{JCAP\ }{\bf #1} (#2) #3}

\def\sss{\scriptscriptstyle}

\begin{document}
\begin{titlepage}
\pagestyle{empty}
\baselineskip=21pt
\rightline{FSU--HEP--050822}
\vskip 0.5in
\begin{center}
{\Huge\sf
\mbox{TeV $\gamma$-rays and the largest masses and}\\[0.4cm] 
\mbox{annihilation cross sections of neutralino dark matter}}
\end{center}
\begin{center}
\vskip 0.6in
{\LARGE\sf Stefano~Profumo}\\
\vskip 0.2in
{\it {Department of Physics, Florida State University, 
Tallahassee, FL 32306, USA}}\\
{E-mail: {\tt profumo@hep.fsu.edu}}\\
\vskip 0.4in
{\bf Abstract}
\end{center}
\baselineskip=18pt \noindent

\noindent Motivated by the interpretation of the recent results on the TeV gamma radiation from the Galactic center, including the new 2004 HESS data, as a by-product of dark matter particles annihilations, we address the question of the largest possible neutralino masses and pair annihilation cross sections in supersymmetric models. Extending the parameter space of minimal models, such as the mSUGRA and the mAMSB scenarios, to general soft SUSY breaking Higgs masses gives access to the largest possible pair annihilation rates, corresponding to resonantly annihilating neutralinos with maximal gaugino-higgsino mixing. Adopting a model-independent approach, we provide analytical and numerical upper limits for the neutralino pair annihilation cross section. A possible loophole is given by the occurrence of non-perturbative electro-weak resonances, a case we also consider here. We then show that a thorough inclusion of QCD effects in gluino (co-)annihilations can, in extreme scenarios, make neutralinos with masses in the hundreds of TeV range, well beyond the s-wave unitarity bound, viable dark matter candidates. Finally, we outline the ranges of neutralino masses and cross sections for models thermally producing a WMAP relic abundance, thus providing reference values for ``best-case'' indirect SUSY dark matter detection rates.

\vfill
\end{titlepage}




\section{Introduction}

The particle physics nature of the non-baryonic and non-luminous
component of the matter budget of the Universe stands today as one of the greatest puzzles in
the understanding of Nature. A few features of this elusive
constituent are relatively well known, including its cosmological
abundance. Structure formation favors a {\em Cold} Dark Matter
(hereafter CDM) particle candidate, {\em i.e.} a particle which
was non-relativistic at the time of its freeze-out from the
thermal bath in the early Universe. Combining a wealth of recent observational
data, including the first year results on cosmic microwave background anisotropies from the WMAP satellite, and assuming a flat $\Lambda$CDM cosmology,
the CDM abundance has been determined to be, at 95\% C.L., \cite{Spergel:2003cb}
\begin{equation}\label{eq:wmap}
\Omega_{\rm\sss CDM}h^2=0.113\pm 0.009
\end{equation}
where $h=0.71\pm0.04$ is the Hubble constant in units of 100 ${\rm km}\ {\rm s}^{-1}\ {\rm Mpc}^{-1}$ .

An attractive class of particle candidates for CDM is that of {\em
weakly interacting massive particles} (WIMPs) (see {\em e.g.}
Ref.~\cite{Bertone:2004pz,Bergstrom:2000pn,Baltz:2004tj} for recent reviews). One of the bonus of the WIMP
scenario is the appealing idea that Dark Matter can be a thermal
leftover from the early Universe, providing a possibly {\em natural} and
{\em fundamental} (in the elementary particle physics sense)
explanation for Eq.~(\ref{eq:wmap}). On the other hand, WIMPs can
be, in principle, detected, either directly, measuring the recoil
energy from elastic scattering on nuclei (see {\em e.g.} \cite{Munoz:2003gx}
for a recent review), or indirectly, looking for the yields of
WIMP pair annihilations. In this latter case, privileged targets
are neutrinos from the center of the Earth or of the Sun \cite{ref:neutrinos}, natural
gravitational traps for WIMPs, gamma rays from spots where dark
matter might have a large density ({\em e.g.} the center of our
Galaxy) \cite{ref:gammarays,Bergstrom:1997fj,Bergstrom:2001jj}, antimatter produced in the galactic halo \cite{ref:positrons,ref:antiprotons,ref:dbar}, or comprehensive multiwavelength analyses which take into account all stable products of WIMP pair annihilations  as well as their secondary yields (see {\em e.g.} the recent analysis of Ref.~\cite{Colafrancesco:2005ji} for the case of the Coma cluster). The estimate
of the fluxes of the mentioned products of WIMP pair
annihilations depends on the product of the thermally
averaged WIMP pair annihilation cross section in the non-relativistic limit, which
will be hereafter indicated as $\langle\sigma v\rangle$, times
(a suitable integral of) the square of the local number density of
WIMPs. The latter quantity is inversely proportional to the WIMP
mass $m_{\rm \sss WIMP}$, at a given Dark Matter density, hence the relevant quantity
for WIMP indirect detection is $\langle\sigma v\rangle/m_{\rm\sss
WIMP}^2$.

Given a fundamental theory which provides a viable WIMP candidate,
one of the theorist's primary goals is to understand the overall
theoretically viable range of a few basic quantities, such as the {\em WIMP
mass}, its {\em scattering cross section off matter} and the
above-mentioned {\em ratio of the pair annihilation cross section
over the WIMP mass squared}. A desirable, complementary
requirement is that the WIMP under consideration {\em thermally}
produces a {\em relic abundance} within the range of
Eq.~(\ref{eq:wmap}). The resulting theoretical information is,
needless to say, crucial, {\em e.g.}, to the design and prospects forecast of WIMP search experiments, or to the interpretation of experimental
results in terms of a WIMP-induced signal.

Among the few well-motivated extensions of the Standard Model (SM)
of particle physics encompassing a WIMP candidate, the option of
low-energy supersymmetry has attracted an extensive, and privileged, investigation
(see Ref.~\cite{Jungman:1995df} for a review). The minimal, $R$-parity
conserving, supersymmetric extension of the SM (MSSM) offers an
ideal WIMP candidate, the lightest neutralino. Contrary to other
frameworks, for instance the DM candidates of universal extra
dimensional scenarios, the viable realizations of the MSSM offer a
wide range of outcomes for the above mentioned DM particle
properties, reflecting the {\em a priori} lack of information on
 the supersymmetry breaking mechanism, and hence on the
supersymmetry breaking lagrangian (see Ref.~\cite{Chung:2003fi} for a review).

Restricting our analysis to the theoretical laboratory of
supersymmetric models featuring a neutralino $\chi$ as the lightest
supersymmetric particle (LSP), we apply here the program of
systematically investigating the resulting {\em maximal mass}
$m_\chi$ and pair annihilation {\em cross sections} $\langle\sigma v\rangle$. As a
concrete, worked-out case study, we assess the possibility of
explaining the recent atmospheric Cherenkov telescopes (ACT) data on the
gamma ray flux from the Galactic center in terms of well defined and motivated neutralino DM setups, a scenario which involves particularly large values for both $m_\chi$ and $\langle\sigma v\rangle$.

In greater detail, we analyze in Sec.~\ref{sec:act} the
top-down interpretation of the Cangaroo-II and of the HESS data on
the high energy gamma rays flux from the Galactic center, nailing down both the
statistically preferred WIMP mass range and the WIMP pair
annihilation cross section for a given central Dark Matter
density. We show, in particular, that the new HESS data from the 2004 campaign highly restrict the class of WIMP candidates which provide statistically viable fits to observations. In Sec. \ref{sec:minimalmodels} we address the question whether
minimal, benchmark, gravity and anomaly mediated supersymmetry breaking 
models are suitable to explain the ACT data, and which are the
regions, on the $(m_\chi,\langle\sigma v\rangle)$ plane, which those
models span. We then point out, in Sec.~\ref{sec:nuhm}, that
minimal extensions of these benchmark models, involving
non-universalities in the Higgs soft supersymmetry breaking masses, allow to largely extend
the maximal neutralino masses compatible with a thermal relic
population of neutralinos making up the CDM. Further, those
minimal extensions feature large annihilation cross
sections, and, remarkably enough, the heaviest neutralinos will most likely be within reach of
future direct DM search experiments. The following
Sec.~\ref{sec:ann} is devoted to a detailed model-independent numerical and
analytical discussion of the maximal neutralino pair annihilation
cross section in the MSSM. The latter is achieved in
correspondence to resonant annihilation processes of maximally higgsino-gaugino mixed neutralinos, with the
possible caveat of special, model dependent values of the
neutralino mass, for which resonant non-perturbative electro-weak effects \cite{Hisano:2004ds} can produce even
larger cross sections. The resulting gamma rays spectral function, in this case, is however shown to be unfit to explain the HESS data. In Sec.~\ref{sec:gluino} we point out that
the inclusion of a non-perturbative QCD treatment of the gluino pair
annihilation cross section can largely affect the relic abundance
of a co-annihilating neutralino LSP. In certain scenarios, the explicit violation of the
$s$-wave unitarity limit, in models with gluino coannihilations,
yields the possibility that neutralinos as heavy as hundreds of
TeV give a thermal relic abundance in the range of
Eq.~(\ref{eq:wmap}). The concluding Sec.~\ref{sec:outlook} gives an
outlook on the maximal cross sections and ``{\em supersymmetric factors}''
$\langle\sigma v\rangle/m_{\chi}^2$ in the MSSM, both for
low-relic density models and for models with neutralino relic
abundances in the WMAP preferred range. A concise summary of the
results presented in this note is provided in the final
Sec.~\ref{sec:conclusions}.



\section{TeV gamma rays from WIMP annihilations: a model-independent analysis}\label{sec:act}

Two ground based ACTs located in the Southern Hemisphere recently 
reported the detection of TeV gamma-rays from the direction of the
Galactic center. The Cangaroo-II telescope observed, during 2001 and 2002, a very soft spectrum,
with a spectral index $\alpha\simeq-4.6$ \cite{Tsuchiya:2004wv}. The HESS
collaboration measured, during Summer 2003, when two of the four telescopes were operational, a substantially
different flux from the Galactic center, featuring a spectral
index $\alpha\simeq-2.2$, and achieving a remarkably better
angular resolution \cite{Aharonian:2004wa}. The results of the 2004 HESS campaign of further
observations of the Galactic center with the complete array of four imaging ACTs were reported this
year in various conferences \cite{hess2004,icrc_rolland,icrc_ripken}, and essentially confirm, with greatly 
improved statistics, the 2003 data. Further observations
with Cangaroo-III will certainly help clarify the situation, which
might be ascribed to the presence of multiple sources not
resolvable by single telescope observations, or which may depend
on peculiar instrumental issues, maybe the same which caused other
reported discrepancies for various different steady sources
observed by both ACTs \cite{canghessdisc}. A further reason could be a time
variability, on a timescale of around one year, of the observed
source; an option which appears highly unlikely, since none of the
experiments detected any significant source variability \cite{Aharonian:2004wa,icrc_rolland}.

A few ``conventional'' astrophysics models have been proposed to
explain the very high energy gamma rays spectrum observed by
ACTs in the center of the Galaxy, ranging from physics involving the central
supermassive black hole \cite{Aharonian:2004wa,Aharonian:2004jr} to the production of gamma rays from the interaction of accelerated protons (possibly injected by the supernova remnant Sgr A East \cite{maeda}) with the ambient matter \cite{Fatuzzo:2003nw,Aharonian:2004wa}. Another option which has been
investigated is that of the pair annihilation of Dark Matter
particles living in the proximity of the central regions of the
Galaxy, giving raise to a gamma-ray continuum from the subsequent
decays of the particles' final state products.

In particular, the latter possibility was pursued in Ref.~\cite{Hooper:2004vp} where the Cangaroo-II and the Whipple data from the VERITAS collaboration \cite{Kosack:2004ri} were analyzed in a model-independent approach using an analytical approximation \cite{Bergstrom:1997fj,Bergstrom:2001jj} to a putative gamma-ray spectrum from a DM particles mainly annihilating into gauge bosons. A correlation between the Dark Matter density in the central region of the Galaxy and the DM particle pair annihilation cross section was also presented, at given values of the DM particle mass. A similar analysis, again based on the same analytical approximation to the gamma-ray spectrum generated by DM pair annihilations, was carried out in Ref.~\cite{Horns:2004bk} for the 2003 HESS data. Although no particular fully motivated particle physics setups were considered in those papers, since minimal supersymmetric models predict neutralino masses typically much lighter than the preferred mass range needed to fit the HESS data, as determined in Ref.~\cite{Horns:2004bk}, a novel, special particle physics scenario was proposed in Ref.~\cite{Hooper:2004fh}. The latter features a stable messenger state with masses in the tens of TeV range as the DM candidate, in the context of an extended version of gauge mediated supersymmetry breaking models involving an additional Higgs singlet. The option of Kaluza Klein DM was finally considered in Ref.~\cite{Bergstrom:2004cy}. The analysis of Ref.~\cite{Bergstrom:2004cy} also included contributions from internal bremsstrahlung. Although the lightest Kaluza-Klein particle (LKP) fails to produce a sufficiently low thermal relic abundance and a large enough LKP pair annihilation cross section (without invoking huge boost factors) at particle masses relevant to fit the full gamma rays energy range spanned by the 2003 HESS data, it was shown that an hypothetical particle with unsuppressed couplings to charged leptons and with increased gauge couplings might give a fairly good fit to the data. In this scenario, however, a central Dark Matter density at least a factor 1000 larger than what predicted by the cuspy Navarro Frank and White profile \cite{nfw} would still be needed. Correlations between the ACT data and the Egret data \cite{egret}, and the possibility of a combined DM annihilation interpretation of the 2003 HESS data and of the Cangaroo-II data were also addressed in Ref.~\cite{Mambrini:2005vk,Fornengo:2004kj}. No specific particle physics models were however proposed there to account for the large masses and annihilation cross sections to be invoked to explain the HESS data. In Ref.~\cite{icrc_ripken} the 2004 HESS data were analyzed in terms of supersymmetric or Kaluza-Klein (KK) DM annihilations. None of these models was found to give satisfactory fits to the 2004 data, and model-dependent limits were set on the WIMPs annihilation cross sections under the assumption of a best-fit power-law background.

In full generality, the continuum gamma rays spectrum generated by the pair annihilation of a WIMP $\chi$ can be cast as \cite{Hooper:2004vp}
\begin{equation}\label{eq:phigamma}
\frac{{\rm d}\Phi_\gamma}{{\rm d}E_\gamma}\simeq5.6\times 10^{-12}{\rm cm}^{-2}{\rm s}^{-1}\left(\sum_f\ {\rm BR}(\chi\chi\rightarrow f)\frac{{\rm d}N^f_\gamma}{{\rm d}E_\gamma}\right)({\mathbf\sigma J})\left(\frac{m_{\chi}}{1\ {\rm TeV}}\right)^{-2}\Delta\Omega,
\end{equation}
where the symbol $f$ refers to any final state of the WIMP pair annihilation process, yielding a gamma rays spectral function (differential number of photons per WIMP annihilation) ${\rm d}N^f_\gamma/{\rm d}E_\gamma$, and where we introduced the quantity $({\mathbf\sigma J})$, defined as
\begin{equation}\label{eq:sigmaj}
({\mathbf\sigma J})\ \equiv\ \left(\frac{\langle\sigma v\rangle}{3\times 10^{-26}{\rm cm}^{3}{\rm s}^{-1}}\right)\times\overline{J(\Delta\Omega)},
\end{equation}
where $\overline{J(\Delta\Omega)}$ is the average, over the solid angle $\Delta\Omega$, of the following line of sight (l.o.s.) integral over the Dark Matter density $\rho_{\rm\sss DM}$
\begin{equation}\label{eq:jpsi}
J(\psi)\equiv\frac{1}{8.5\ {\rm kpc}}\left(\frac{1}{0.3\ {\rm GeV}/{\rm cm}^3}\right)^2\int_{\rm l.o.s.}{\rm d}l(\psi)\rho_{\rm\sss DM}^2(l).
\end{equation}
In the following analysis, we assume a solid angle $\Delta\Omega\sim5\times10^{-5}$ for the Cangaroo-II telescope, corresponding to an angular resolution around $0.2^{\rm\sss 0}$, while we consider, for HESS, an angular resolution of $5.8^\prime$, corresponding to the instrumental 50\% containment radius\footnote{This yields a further factor 0.5 in Eq.~(\ref{eq:phigamma}) for the case of HESS.} \cite{Aharonian:2004wa}. The latter superior angular resolution will give rise to slightly larger values of $\overline{J(\Delta\Omega)}$, particularly in the relevant case of cuspy DM profiles.
  
In this Section we take the opportunity to analyze the recently reported 2004 HESS data in terms of DM annihilations, adopting a fully model-independent strategy, which we also apply, for comparison, to the HESS 2003 and to the Cangaroo-II results. The purpose of this study is two-fold: first, we shall assess the preferred range for the DM masses and pair annihilation cross sections in the top-down interpretation of the ACTs data, in the interest of, and as a motivation to, the remainder of this paper; second, we wish to extend and generalize the above mentioned analyses of ACTs data, privileging a fully numerical and model-independent approach. In this respect, instead of analytically approximating the gamma rays spectrum generated by DM pair annihilations, we use the Monte-Carlo simulations results from the {\tt Pythia} code \cite{pythia} for the gamma rays spectral functions, as implemented in the {\tt DarkSUSY} package \cite{Gondolo:2004sc}. Further, we numerically determine the ``{\em best spectral functions}'', {\em i.e.} the set of branching ratios minimizing the $\chi^2$ of a given ACT data set fit, hence providing the most conservative confidence level regions for the DM particle mass and annihilation cross section.

As a first step, we want to pinpoint the pair annihilating DM mass range and final state branching ratios pattern favored by the HESS data. Given a set of branching ratios $\{a_f\}$, such that $\sum_f a_f=1$, and a WIMP mass $m_{\chi}$, one can analytically determine the value of the quantity $(\sigma J)$, defined in Eq.~(\ref{eq:sigmaj}), providing the lowest $\chi^2$ in the fit to a certain experimental data set $\{i\}$ with data points $(E_\gamma^i,\Phi^i_\gamma\pm\Delta\Phi^i_\gamma))$, to be

{\small
\begin{equation}
\label{eq:bestchi2}
(\sigma J)\equiv\left(\sum_{i\in\{i\}}\frac{\sum_f a_f\frac{{\rm d}N^f_\gamma}{{\rm d}E_\gamma}(E_\gamma^i)\cdot\Phi^i_\gamma}{\left(\Delta\Phi^i_\gamma\right)^2}\right)\times\left(\sum_{i\in\{i\}}5.6\times 10^{-12}{\rm cm}^{-2}{\rm s}^{-1}\left(\frac{m_\chi}{1\ {\rm TeV}}\right)^{-2}\Delta\Omega\left(\sum_f a_f\frac{{\rm d}N^f_\gamma}{{\rm d}E_\gamma}(E_\gamma^i)\right)^2\left(\Delta\Phi^i_\gamma\right)^{-2}\right)^{-1}.
\end{equation}
}
As a side remark, in the computation of the $\chi^2$, as well as in Eq.~(\ref{eq:bestchi2}), we take into account the finite energy resolution of the HESS detector, which is $\sim15\%$ \cite{icrc_rolland}, and consider, for every data point, the average WIMP induced gamma rays flux in each resulting finite energy interval. Eq.~(\ref{eq:bestchi2}) allows to determine the minimal $\chi^2$ for a given WIMP model (with the branching ratios set $\{a_f\}$), at each annihilating particle mass. In order to find the absolute $\chi^2$ minimum, one needs to find the ``{\em best}''  $\{a_f\}$ set. To this extent, we applied a Monte Carlo technique: we started from a random configuration $\{a_f\}_0$, computing, through equation (\ref{eq:bestchi2}), the associated minimal $\chi^2_0$. We then generated a new configuration $\{a_f\}_1$, varying one of the coefficients, and re-normalizing the whole set so that $\sum_f a_f=1$ (the size of the variation was chosen to optimize the convergence to the minimum). The new set $\{a_f\}_1$ was then accepted with probability 1 if the resulting $\chi^2_1<\chi^2_0$, and with probability $\exp(-(\chi^2_1-\chi^2_0)/2)$ if $\chi^2_1>\chi^2_0$ (in order to avoid local minima in the parameter space). The whole procedure was then re-iterated, keeping track of the absolute lowest $\chi^2$ value reached along the resulting ``Markov chain''. The associated branching ratio set $\{a_f\}_{\rm min}$ was eventually selected as providing the ``{\em best spectral function}'' at that particular WIMP mass.

\begin{figure}[!t]
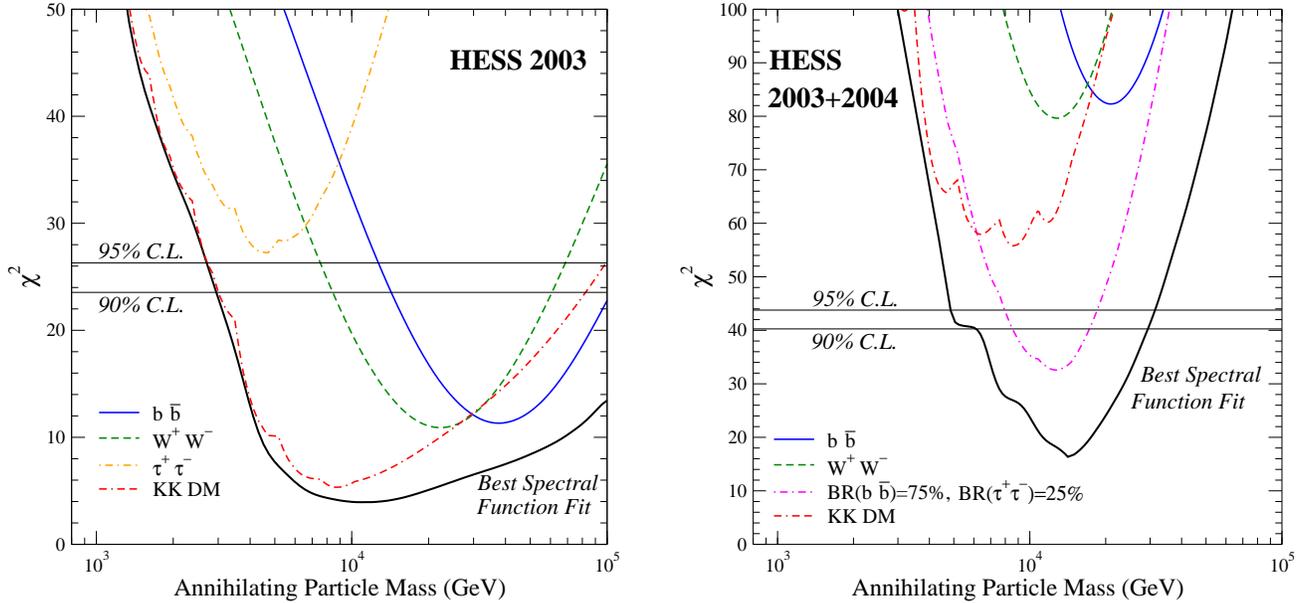

\begin{center}
\mbox{\epsfig{file=chi2_hess_2003.eps,height=8.cm}\quad\quad\epsfig{file=chi2_hess_all.eps,height=8.cm}}
\end{center}
\caption{\it\small  The minimal $\chi^2$ resulting from fits to the 2003 HESS data (left panel), and to the full 2003+2004 HESS data set (right panel) on the gamma-rays flux from the Galactic Center, as a function of the annihilating Dark Matter particle mass, for different final state channels (see the text for details), including, as a thick black line, the ``{\em best spectral function}'' case. The two horizontal lines in each panel indicate the 90\% and 95\% C.L. exclusion limits corresponding to the set of data under consideration.
}
\label{fig:chi2}
\end{figure}
Besides this fully model-independent approach, we also picked a few representative spectral functions. In particular, we took the case of a DM particle annihilating with 100\% branching ratio into a given final state. Based on the spectral features of the gamma ray spectrum (see {\em e.g.} Ref.~\cite{Fornengo:2004kj,Colafrancesco:2005ji}), we singled out three significant final states, namely (ordering from the {\em softer} to the {\em harder} spectrum) a heavy quark-antiquark pair (for definiteness, $b\bar b$), a pair of massive gauge bosons ($W^+W^-$), and a pair of taus ($\tau^+\tau^-$), including in the latter case the possibility of a photon in the final state. We also addressed the case of Kaluza-Klein (KK) DM, in the context of Universal Extra Dimensions. We included in our analysis the internal bremsstrahlung effect discussed in Ref.~\cite{Bergstrom:2004cy}, and we made use the branching ratios obtained in Ref.~\cite{Servant:2002aq}. The resulting spectrum is quite hard, and significantly different from those mentioned above, mainly due to the helicity unsuppressed leptonic final states contributing to the process $B^{(1)}B^{(1)}\rightarrow l^+l^-\gamma$. 

We show our results, for the three final states $b\bar b$, $W^+W^-$, $\tau^+\tau^-$, for the case of KK DM and for the ``{\em best spectral function}'', in Fig.~\ref{fig:chi2}, for the the 2003 HESS data (left panel) and for the full HESS data set (2003+2004 data, right panel). We notice that WIMPs purely annihilating in $b\bar b$ or in $W^+W^-$, and KK DM particles give, in different ranges of masses, statistically acceptable fits to the 2003 HESS data: this means, in particular, that {\em the spectral features of the annihilating particle were not tightly constrained by the 2003 HESS data}. Including the new 2004 HESS data, instead, statistically {\em rules out}, with a high confidence level, both the cases of a WIMP purely annihilating into gauge bosons (a final state which applies to many supersymmetric models, including the case of pure higgsinos and winos) or quark pairs, and the KK DM scenario \cite{Bergstrom:2004cy}. 

In order to improve the fit, we resorted to a mixed final state composed by a branching ratio $x$ into $\tau^+\tau^-$ and $1-x$ into $b\bar b$. The resulting spectral function features a harder spectra at \mbox{$E_\gamma/m_{\chi}\rightarrow 1$}, and improves the fit at the larger energies probed by the 2004 HESS campaign. Further, this particular final state is well motivated, in the framework of supersymmetric DM. At large $\tan\beta$, bino-like neutralinos tend to have a final state pattern as that considered above, with a naively estimated relative weight, on the basis of color factors only, and neglecting the fermions and sfermions mass effects, $x\approx0.25=1/(3+1)$. In this case, a region allowed at 90\% C.L. appears for the mixed final state case. 

We also considered the internal bremsstrahlung of $W$ pairs discussed in Ref.~\cite{Bergstrom:2005ss}, which gives a further hard component to the spectral function of the $W^+W^-$ final state at \mbox{$E_\gamma/m_{\chi}\rightarrow 1$}. The resulting minimal $\chi^2$ for a WIMP annihilating with branching ratio 1 into charged gauge bosons ({\em e.g.} a wino-like neutralino) is typically lower than what shown in Fig.~\ref{fig:chi2}, but it is a rapidly oscillating function of the mass. This simply depends on where the harder end of the spectral function (at $E_\gamma\lesssim m_\chi$) is located with respect to the HESS data points. In any case, we find that the minimal $\chi^2$'s per degree of freedom are always larger than $\approx50/30$, ruling out this scenario at 99\% C.L..

Resorting, finally, to the ``{\em best spectral functions}'', the relevant, most conservative mass range allowed at 90\% C.L. reads
\begin{equation}
\label{eq:hess2004mass}
6\ <\ m_\chi/{\rm TeV}\ <\ 30.
\end{equation}
The WIMP models providing the best fit to the HESS data set are found to feature spectral functions given by suitable mixtures of soft and hard channels. On average, we find that the total branching ratio into quark pairs (``soft channels'') lies between $40\%\div70\%$, that into gauge bosons is always less than 20\% and that into ``hard'' channels (e.g. charged lepton pairs) lies between $20\%\div40\%$, the fluctuations in these figures being rather significant, depending on the WIMP mass.

\begin{figure}[!t]
\begin{center}
\epsfig{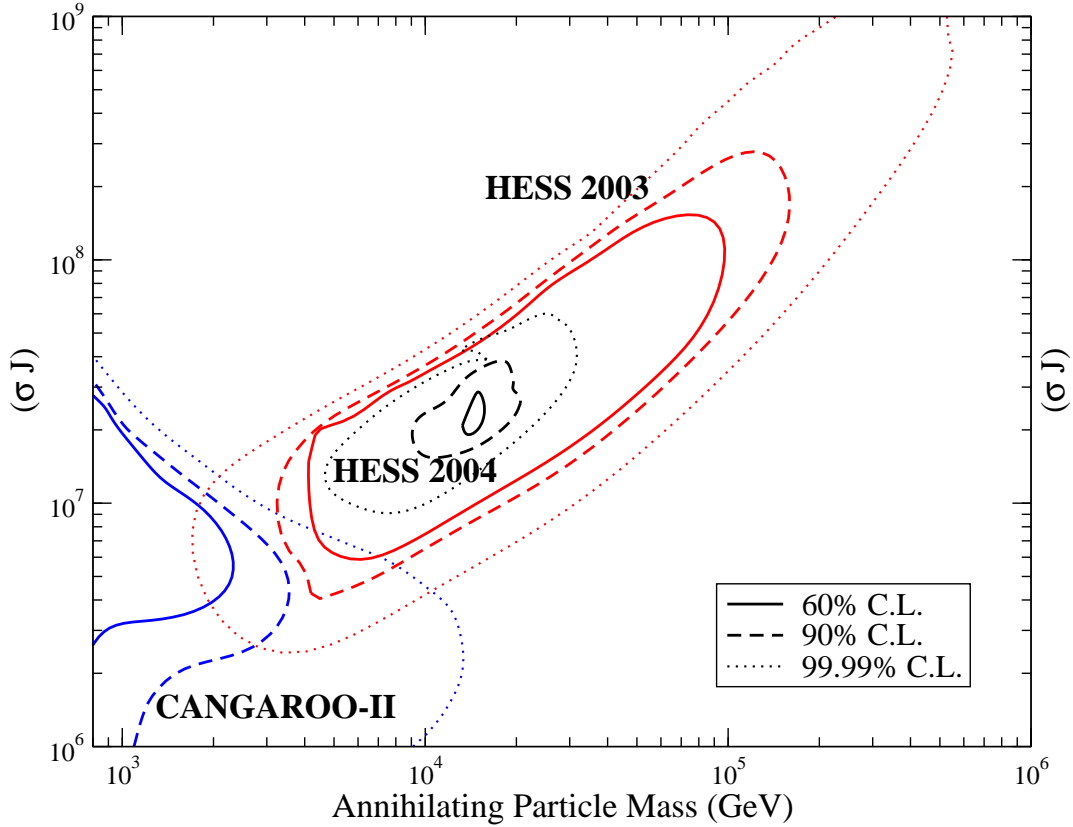}
\end{center}
\caption{\it\small  Iso-confidence-level contours of ``{\em best spectral functions}'' fits to the Cangaroo-II and to the 2003 and 2004 HESS data, in the plane defined by the annihilating particle mass and by the quantity $(\sigma J)$, defined in Eq.~(\protect{\ref{eq:sigmaj}}).
}
\label{fig:cntr}
\end{figure}
To have a more exhaustive picture, we show in Fig.~\ref{fig:cntr} the most conservative ({\em i.e.} computed in the best spectral function approach) C.L. regions for the Cangaroo-II and HESS data (we separately analyze here the 2003 and 2004 data), on the full plane spanned by the annihilating WIMP mass versus the quantity $(\sigma J)$ defined in Eq.~(\ref{eq:sigmaj}). We adopt here, at every $m_\chi$, the same procedure of Monte Carlo $\chi^2$ minimization described above, dropping, naturally, the use of Eq.~(\ref{eq:bestchi2}). As discussed above, the 2004 HESS data significantly shrank the allowed region, which, quite remarkably, is fully compatible with the best fit region of the 2003 data. The different spectral indexes found by the Cangaroo-II and HESS collaborations explain the unsimilar shape of the favored $(m_\chi,(\sigma J))$ regions (particularly in the way the best $(\sigma J)$ values vary with $m_\chi$), which are clearly incompatible among each other.

\begin{figure}[!t]
\begin{center}
\epsfig{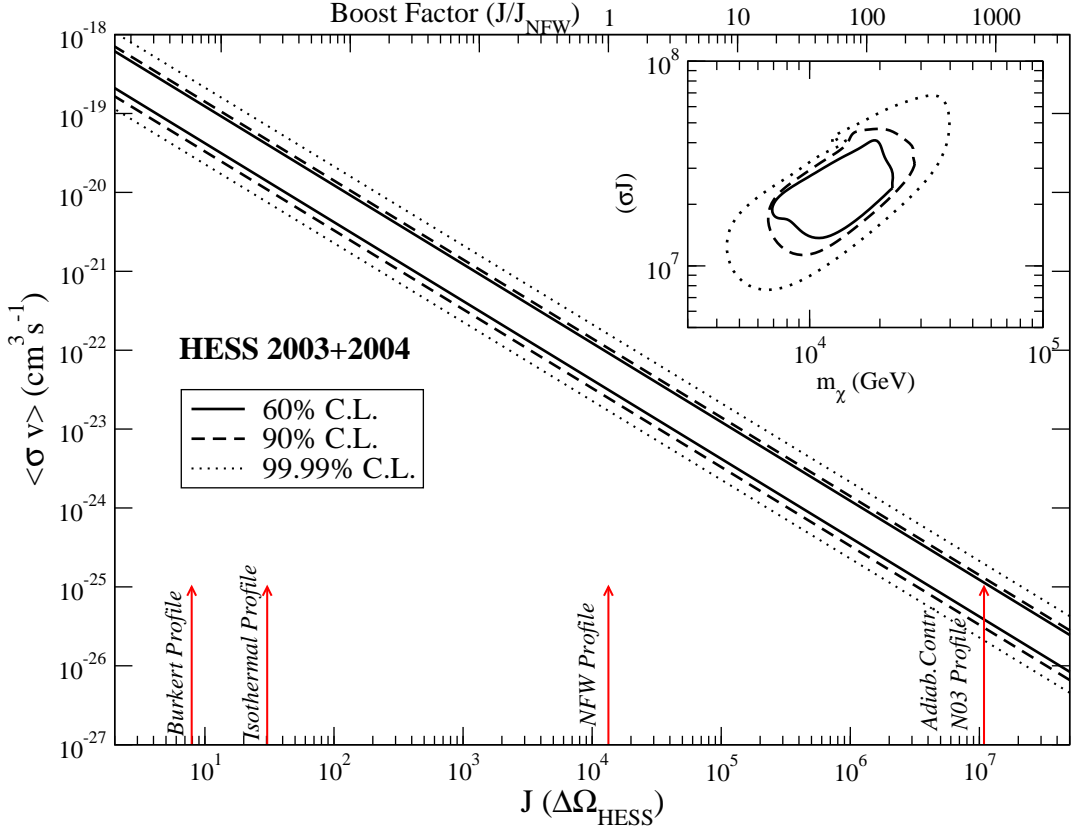}
\end{center}
\caption{\it\small  Confidence level contours in the $(\langle\sigma v\rangle,J)$ plane, for the full HESS data set. The upper $x$-axis indicates the ``boost factor'', with respect to the NFW halo profile, {\em i.e.} $J/J_{NFW}$. The red arrows indicate the values of $J$ for a sample of viable Milky Ways halo profiles. In the upper right panel we show the confidence level regions on the $(m_{\chi},(\sigma J))$ plane for the 2003+2004 HESS data, in the ``{\em best spectral functions}'' approach.
}
\label{fig:sigmaj}
\end{figure}
The displayed confidence level contours determine the ranges of $(\sigma J)$ allowed at a given C.L., and hence enable to nail down the relevant intervals on the physical $(\overline{J(\Delta\Omega)},\langle\sigma v\rangle)$ plane. This is done in Fig.~\ref{fig:sigmaj}, where we jointly analyze the full 2003+2004 HESS data set (the analogous of Fig.~\ref{fig:cntr} is shown in the small panel in the upper right). The red arrows on the $x$-axis indicate the value of the quantity $\overline{J(\Delta\Omega)}$, defined in Eq.~(\ref{eq:jpsi}), computed for the relevant solid angle $\Delta\Omega_{\rm \sss HESS}$, for four different halo model profiles, ranging from the cored Burkert profile \cite{burkert} to the extremely cuspy adiabatic contraction \cite{blumental} of the N03 profile \cite{n03} (ACN03 below) (for the definition of the halo models under consideration here see also Ref.~\cite{pierohalos} and Ref.~\cite{Gondolo:2004sc}). The upper $x$-axis indicates the so-called ``{\em boost factor}'', defined as the deviation of $\overline{J(\Delta\Omega)}$ from its value for the Navarro Frenk and White (NFW) profile, {\em i.e.} $\overline{J(\Delta\Omega)}/\overline{J_{\rm NFW}\Delta\Omega}$ \cite{Bergstrom:2004cy}. At 90\% C.L., and focussing on the two cuspiest profiles under consideration, respectively the NFW profile and the ACN03 profile, we find that the relevant intervals for the DM annihilation cross section, in units of ${\rm cm}^3{\rm s}^{-1}$ and considering the full 2003+2004 HESS data set, read:

\begin{equation}
\label{eq:hess_sv} 2.5\times10^{-23}<\langle\sigma v\rangle< 1.1\times10^{-22}\quad ({\rm NFW}) \quad 3.0\times10^{-26} <\langle\sigma v\rangle< 1.3\times10^{-25} \quad ({\rm ACN03})
\end{equation}

Assuming the naive relation between a WIMP pair annihilation cross section and its relic abundance 
\begin{equation}\label{eq:naiverelation}
\langle\sigma v\rangle\approx 3\times10^{-27}\ {\rm cm}^3{\rm s}^{-1}/\Omega_\chi h^2, 
\end{equation}
one would conclude, from the above listed ranges, that a WIMP providing the WMAP relic abundance indicated in Eq.~(\ref{eq:wmap}) could yield a large enough gamma ray flux to explain the ACT data {\em only if an extremely cuspy profile}, or a {\em boost factor of the order of $10^3$} is hypothesized for the DM density in the galactic center\footnote{If $\Omega_\chi<\Omega_{\rm CDM}$, and the dark matter is made up of other components besides the neutralino, the neutralino number density squared, relevant for the photon fluxes, would be naturally rescaled by a factor $(\Omega_chi/\Omega_{\rm CDM})^2$, and the needed cross sections or boost factors to explain the ACT data would be even larger.}. We will show below that the above mentioned relation (\ref{eq:naiverelation}) can badly fail, and that not only can very massive ($\mathcal{O}(10\div100)$ TeV) neutralinos produce a relic abundance in the range of  Eq.~(\ref{eq:wmap}), but also that, even at those large masses, supersymmetric models allow for $\langle\sigma v\rangle$ values implying significantly less cuspy DM profiles than what the estimate above would suggest. 


\section{Gravity and anomaly mediated SUSY breaking models}\label{sec:minimal}

The phenomenological investigation of supersymmetric extensions of the Standard Model faces the ineluctable necessity of handling a huge parameter space, a consequence of our ignorance of the mechanism of supersymmetry breaking (see Ref.~\cite{Chung:2003fi} for a recent review of the soft supersymmetry breaking lagrangian). A way out of this conundrum is to assume that a particular physical process dominates the coupling of the hidden sector, where supersymmetry is broken, to the visible sector. A (partial) list of such processes includes gravity, (super-Weyl) anomaly and gauge mediated supersymmetry breaking models \cite{Chung:2003fi,rev}. A set of further assumptions, motivated on theoretical or phenomenological grounds, has lead to the definition of a few {\em minimal} frameworks, featuring a manageable parameter space (see {\em e.g.} Ref.~\cite{Baer:2000gf}). The question of how well these minimal setups faithfully reproduce the phenomenological features of the general MSSM \cite{Profumo:2004at}, or of the subset of MSSM models sharing the same supersymmetry breaking mechanism, is both long-standing and, in many respects, not totally understood. However, in view of the great wealth of phenomenological studies focusing on minimal models, it is certainly worthwhile to consider them as {\em benchmark} setups, and to start our present investigation from them.

A very popular framework which assumes a gravity mediated supersymmetry breaking scenario is that of {\em minimal supergravity} (hereafter mSUGRA, see Ref.~\cite{msugra}). Universal scalar and gaugino soft breaking masses (respectively $m_0$ and $M_{1/2}$) plus a common trilinear scalar coupling ($A_0$), all of them defined at the grand unification (GUT) scale, and the requirement of successful radiative electro-weak symmetry breaking (REWSB) reduce the parameter space of mSUGRA to four continuous parameters plus one sign, {\em viz.},
\begin{equation}\label{eq:msugraps}
m_0,\ M_{1/2},\ A_0,\ \tan\beta,\  {\rm sign}(\mu)
\end{equation}
where $\tan\beta$ indicates the ratio of the Higgs field vacuum expectation values and ${\rm sign}(\mu)$ stands for the sign of the supersymmetric $\mu$ term. 

The minimal anomaly mediated supersymmetry breaking (mAMSB) scenario is motivated by the possible dominance of supersymmetry breaking contributions originating in the super-Weyl anomaly, which are always present when supersymmetry is broken \cite{Giudice:1998xp}. When the supersymmetry breaking and the visible sector reside on different branes, sufficiently separated in a higher dimensional space, gravity contributions can in fact be strongly suppressed \cite{Randall:1998uk}. In this case, the resulting soft parameters are {\em UV insensitive}, and can be expressed in terms of low energy entries, such as the Yukawa and gauge couplings and the gravitino mass, $m_{3/2}$. For instance, the gaugino spectrum is given by
\begin{equation}
M_i=\frac{\beta_{g_i}}{g_i}m_{3/2},
\end{equation}
where $\beta_{g_i}$ indicate the beta functions of the $g_i$ coupling, $i=1,2,3$. This yields a specific relationship in the low energy gaugino mass spectrum, namely
\begin{equation}
M_1:M_2:M_3=2.8:1:7.1,
\end{equation}
which reverse the mSUGRA hierarchy between the bino and wino mass terms ($M_2/M_1\sim 2$). A further universal scalar mass $m_0$ is postulated to cure tachyonic sfermions, and radiative electroweak symmetry breaking is required, thus yielding an overall parameter space consisting of the following set \cite{Gherghetta:1999sw,Feng:1999hg}:
\begin{equation}\label{eq:mamsbps}
\tan\beta,\  m_{3/2},\ m_0,\ {\rm sign}(\mu).
\end{equation} 

\subsection{Maximal neutralino masses in Minimal models: mSUGRA \& mAMSB}\label{sec:minimalmodels}

{\em Can minimal, benchmark supersymmetric models provide adequate particle physics setups to explain the ACT data as analyzed in Sec.~\ref{sec:act}}? As a first step, we need to assess whether the maximal neutralino masses compatible with the observed DM abundance are large enough to fall within the ranges indicated in Fig.~\ref{fig:chi2}. Secondly, those large neutralino mass models must feature a neutralino pair annihilation cross section compatible with what we nailed down in Sec.~\ref{sec:act}.

The composition of the lightest neutralino in mSUGRA is dictated by the renormalization group equations (RGE) of the gaugino masses from the GUT down to the electro-weak scale, giving, at that latter scale, the approximate ratios $M_2/M_1\sim2$ and $M_3/M_1\sim6$, and by the absolute value of $\mu$, as determined by the REWSB conditions. Since in most of the mSUGRA parameter space $|\mu|\gg M_1$, the lightest neutralino is often found to be an almost purely bino-like state. Unless specific mechanisms (which will be shortly reviewed below) are in place, pure binos tend to have a thermal relic abundance well above the range of Eq.~(\ref{eq:wmap}). As a consequence, the parameter space of mSUGRA compatible with a viable DM candidate is limited to a handful of special regions, which we list below:
\begin{enumerate}
\item A {\em bulk} region at low values of $m_0$ and $M_{1/2}$, where the $t$-channel sfermions exchange channel is enhanced by relatively light sfermions \cite{bulk};
\item A {\em stau coannihilation} region, in the region at low $m_0$ where $m_{\widetilde\tau_1}\gtrsim m_{\chi}$ \cite{stau};
\item A resonant annihilation channel through $s$-channel heavy Higgses ({\em funnel} region ), occurring at large $\tan\beta$, when $m_\chi\simeq m_A/2$, $A$ being the CP-odd MSSM Higgs \cite{Afunnel};
\item The {\em hyperbolic branch/focus point} (HB/FP) region, where large values of $m_0$ drive the $\mu$ term to low values through REWSB, enhancing the higgsino fraction of the lightest neutralino \cite{hb_fp,Baer:2005ky};
\item A {\em stop coannihilation} region, at large values of $A_0$ and low values of $m_0$ and $M_{1/2}$ \cite{Boehm:1999bj,stop,Edsjo:2003us};
\item A {\em light Higgs resonant annihilation} corridor, occurring when $m_\chi\simeq m_h/2$ \cite{drees_h}.
\end{enumerate}

It is not difficult to understand that the largest neutralino masses cannot occur in regions 1., 5. and 6., where the universal gaugino mass $M_{1/2}$ is bound to assume low values, well below the TeV (and hence the neutralino mass, which cannot be larger, in those regions, than $\approx0.4\times M_{1/2}$). On the other hand, the maximal neutralino mass in the remaining regions can be determined by ({\em a.}) setting the relic abundance to its maximal value in Eq.~(\ref{eq:wmap}) and ({\em b.}) maximizing the efficiency of the relic density suppression mechanism in place. In particular, this amounts to pick points at $m_{\widetilde\tau_1}=m_{\chi}$ in region 2., points at $m_\chi=m_A/2$ in region 3.\footnote{Since the freeze-out process occurs at finite temperature, the minimal relic abundance in the funnel region strictly occurs at $m_\chi\neq m_A/2$; however, since the induced variation in the maximal mass is small, and since we want to maximize also the neutralino pair annihilation cross section at $T=0$, we take here the simpler condition $m_\chi=m_A/2$.}, and pure higgsinos in region 4. \cite{hb_fp,Baer:2005ky}. 

\begin{figure}[!t]
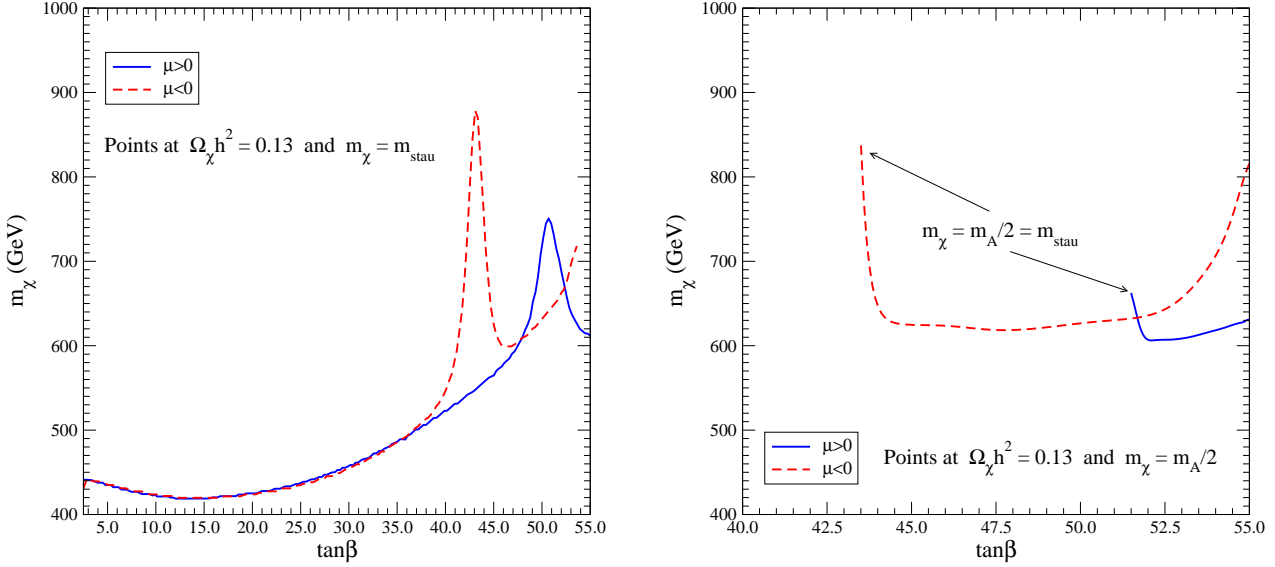

\begin{center}
\hspace*{-0.7cm}\mbox{\epsfig{file=maxmx_coan.eps,height=7.5cm}\quad\quad\epsfig{file=maxmx_funn.eps,height=7.5cm}}
\end{center}
\caption{\it\small  (Left panel): Points, in the mSUGRA parameter space at $A_0=0$ and at $m_{\chi}=m_{\widetilde\tau_1}$, giving $\Omega_\chi h^2=0.13$, {\em i.e.} the largest possible neutralino masses in the stau coannihilation region, for positive (solid blue line) and negative (dashed red line) $\mu$. (Right panel):  Points, in the mSUGRA parameter space at $A_0=0$ and at $m_{\chi}=m_{A}/2$, giving $\Omega_\chi h^2=0.13$, {\em i.e.} the largest possible neutralino masses in the funnel region, for positive (solid blue line) and negative (dashed red line) $\mu$.
}
\label{fig:maxmsmsugra}
\end{figure}
We show in Fig.~\ref{fig:maxmsmsugra} the resulting maximal neutralino masses in region 2. (left panel) and region 3. (right panel), for positive (solid blue line) and negative (red dashed lines) values of ${\rm sign}(\mu)$. The dashed line in the left panel terminates where REWSB is no longer possible, while the lines in the right panel start where $m_\chi=m_{\widetilde\tau_1}$. The peaks in the left panel correspond to the overlap of stau coannihilations and of resonant annihilations, and give the largest neutralino masses in these two regions, $m_\chi\lesssim 900$ GeV. In this analysis, we set $A_0=0$; a non-zero value for the trilinear scalar coupling affects the left-right mixing in the lightest stau, and hence the neutralino relic abundance. We checked that this effect amounts to a factor well within $\pm10\%$, with larger relic abundances at large and {\em positive} $A_0$ and smaller relic abundances at large and {\em negative} $A_0$. Since the relic abundance of the degenerate neutralino-stau system scales as $m_\chi^2$, taking into account the effect of $A_0\neq0$ in Fig.~\ref{fig:maxmsmsugra} amounts to allow an uncertainty factor of $\pm\mathcal O(5\%)$. 

As a side comment, the points in Fig.~\ref{fig:maxmsmsugra}, left, where $m_\chi\simeq m_{\widetilde\tau_1}$, have been shown, in Ref.~\cite{Profumo:2004qt}, to play a special role in the suppression of unwanted small scale (sub-galactic) structures, in virtue of the effects of a long-lived stau in structure formation. Regarding region 4., the reader is referred to the model-independent analysis of Ref.~\cite{Baer:2005ky}; the maximal neutralino mass in the focus point region ranges between 1150 and 1170 GeV, slightly varying with $\tan\beta$. As a by-product, we also conclude that {\em within mSUGRA, the largest neutralino mass compatible with a thermal relic abundance consistent with the WMAP result is less than 1.2 TeV}.

In the case of the mAMSB model, the parameter space is much more homogeneous \cite{Randall:1998uk,Gherghetta:1999sw,Feng:1999hg}, and the lightest neutralino is almost everywhere wino-like to a high degree of ``purity''. Since winos efficiently annihilate into gauge bosons, and undergo coannihilations with the lightest chargino, with a small spread in the relative mass splitting, provided the lightest neutralino is the LSP its relic abundance smoothly depends on its mass, according to the functional relation \cite{splitsusy}
\begin{equation}\label{eq:winoomega}
\Omega_\chi h^2\ \simeq\ c\cdot\left(\frac{m_\chi}{1\ {\rm TeV}}\right)^\gamma, \qquad\quad 0.0225<c<0.0255, \quad \gamma\simeq1.90\div1.92
\end{equation}
the spread in the coefficients being motivated by the details of the model particles spectrum. As a consequence, it is easy to figure out the range of neutralino masses falling within the WMAP range. A numerical scan of the mAMSB parameter space indicates that the relic abundance of mAMSB neutralinos falls within the range of Eq.~(\ref{eq:wmap}) for neutralino masses $2.0\ {\rm TeV}\lesssim m_\chi\lesssim 2.6\ {\rm TeV}$.

\begin{figure}[!t]
\begin{center}
\hspace*{-1cm}\epsfig{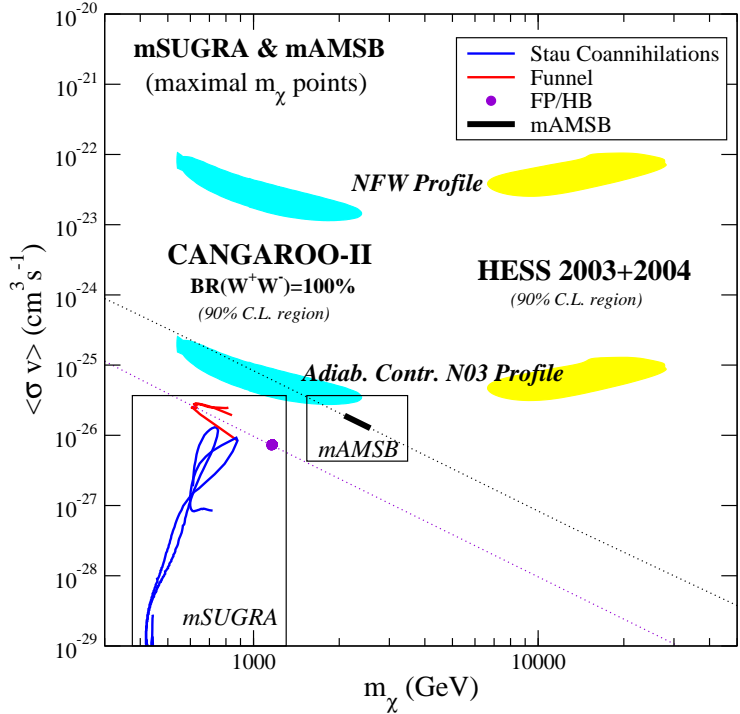}
\end{center}
\caption{\it\small  The pair annihilation cross sections of the largest mass neutralinos in the mSUGRA and mAMSB models giving a WMAP neutralino relic density. The shaded areas correspond to the 90\% C.L. contours for the Cangaroo-II data (left, gauge bosons final state case) and for the HESS data (right, best spectral function case, the gauge boson final state being ruled out, see Fig.~\ref{fig:chi2}, right), for a NFW profile (upper shaded areas) and for the adiabatically contracted N03 profile (lower shaded areas). The upper (lower) dotted lines correspond to extrapolations of the cross section of mAMSB (HB/FP region of mSUGRA) neutralinos giving a relic abundance outside the 2-$\sigma$ WMAP range.
}
\label{fig:mx_sv_all}
\end{figure}
We can already draw, at this point, the conclusion that minimal models do not give large enough neutralino masses to explain the HESS data (see (\ref{eq:hess2004mass})). In order to understand whether these models feature a large enough pair annihilation cross section to explain, at least, the Cangaroo-II data, we map the mSUGRA regions corresponding to the maximal neutralino masses, and the mAMSB points with WMAP relic abundances, on the $(m_\chi,\langle\sigma v\rangle)$ plane. The neutralino pair annihilation cross sections have been computed, here and in the remainder of this paper, with the {\tt DarkSUSY} package \cite{Gondolo:2004sc}. We shade in light blue the 90\% C.L. contours corresponding to fits to the Cangaroo-II data with a 100\% branching ratio into gauge bosons pairs, the case relevant for the largest neutralino masses under consideration here. We also include, for completeness, the best spectral function 90\% C.L. region corresponding to the (2003+2004) HESS data. The two upper contours refer to the NFW profile, while the lower ones to the ACN03 profile. We also extrapolated the mAMSB (pure wino) and FP (pure higgsino) lines to models featuring a lower (left portions of the dotted lines) and higher (right) thermal relic abundance. We recall that low thermal relic density models can provide all the required DM in the context of non-standard cosmological scenarios (including a quintessential energy density term \cite{quint}, Brans-Dicke-Jordan theories \cite{Kamionkowski:1990ni,bdj} and anisotropic cosmologies \cite{Kamionkowski:1990ni,Profumo:2004ex}) or of non-thermal production \cite{nonth}. We henceforth conclude that 
\begin{itemize}
\item {\em the mSUGRA model is unfit to explain ACT data}, and that
\item {\em the mAMSB model can explain the Cangaroo-II data only, and with (1) an extremely cuspy DM profile as well as (2) some kind of cosmological relic density enhancement mechanism}.
\end{itemize}

\subsection{Minimal models with non-universal Higgs masses}\label{sec:nuhm}

A common and reasonable approach to fill the gap between {\em minimal} models and the {\em general} MSSM is to relax some of the assumptions of universality, possibly following GUT-motivated guidelines. In the present framework, relaxing universality in the sfermions sector \cite{nonuniversalscalar} would not greatly help, since while larger neutralino masses could be made compatible with the DM abundance, the pair annihilation cross section would be hardly affected. The two remaining options concern the {\em gaugino} and the {\em Higgs} sector of soft supersymmetry breaking masses. Both possibilities are well motivated by supersymmetric GUTs, and have been widely investigated in the literature \cite{nugm,nuhm}. We pick here the option of non-universality in the Higgs sector, which has been recently analyzed in great detail in Ref.~\cite{Baer:2005bu} for the case of mSUGRA. We consider here, for the first time, general soft supersymmetry breaking Higgs mass extensions to the mAMSB as well. It will be shown, in the following Sec.~\ref{sec:ann}, that resorting to what we will hereafter dub as {\em non-universal Higgs masses} (NUHM) models actually allows to access the largest possible neutralino pair annihilation cross sections of the general MSSM.

Considering general soft breaking mass terms $m_{H_u}$ and $m_{H_d}$ for the Higgs doublet superfields $\hat H_u$ and $\hat H_d$ amounts, through the conditions of REWSB, to extend the minimal model parameter spaces of Eq.~(\ref{eq:msugraps}) and (\ref{eq:mamsbps}) to the extra weak-scale parameters $\mu$ and $m_A$ \cite{nuhm}. Stretching the neutralino masses to the largest possible values compatible with a WMAP thermal relic abundance squeezes the NUHM parameter space to the regions where resonant neutralino annihilation occurs ($m_\chi\approx m_A/2$). Further, since the neutralino couplings to the Higgses scale as the product of the gaugino and of the higgsino fractions, the lowest relic abundances are found in the regions where a maximal higgsino-gaugino mixing occurs.

\begin{figure}[!t]
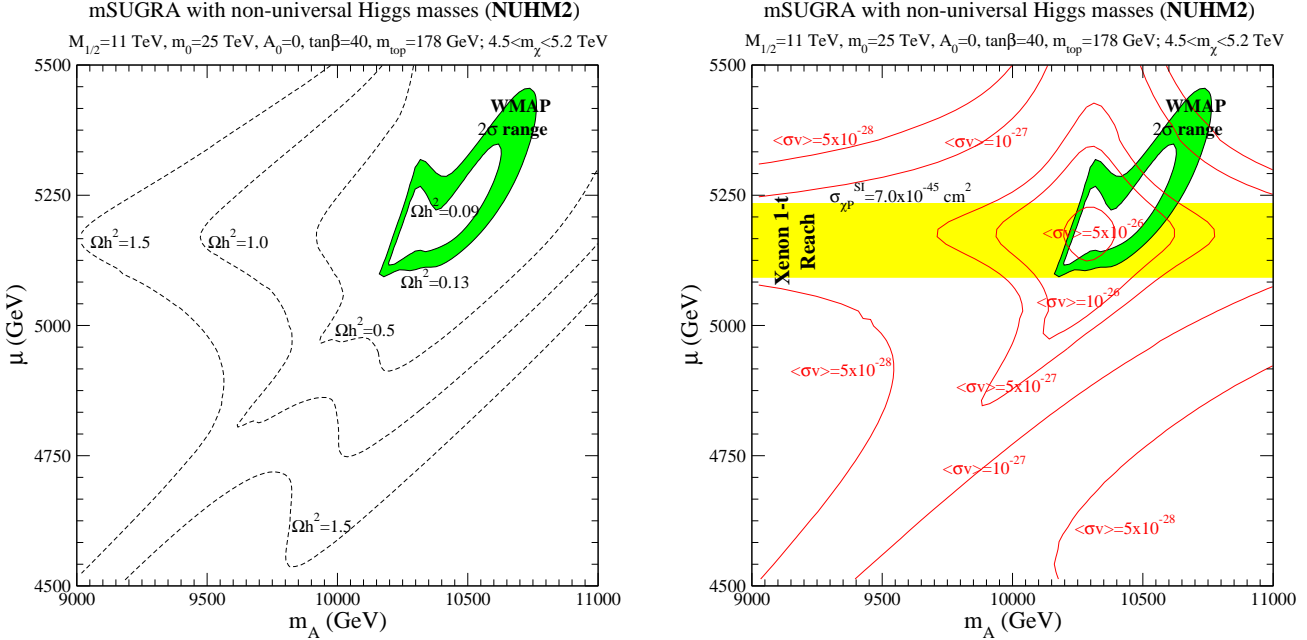

\begin{center}
\hspace*{-0.7cm}\mbox{\epsfig{file=mamu_oh2.eps,height=8.5cm}\quad\quad\epsfig{file=mamu_sv.eps,height=8.5cm}}
\end{center}
\caption{\it\small  (Left): Curves of iso-relic abundance for neutralinos in the mSUGRA model with non-universal Higgs masses, in the $(m_A,\mu)$ plane. The green shaded area corresponds to the 2-$\sigma$ WMAP range. (Right): Curves of iso-cross section for neutralinos in the mSUGRA model with non-universal Higgs masses, in the $(m_A,\mu)$ plane. The green shaded area corresponds to the 2-$\sigma$ WMAP range, while the yellow shaded area indicates the reach of stage-3, ton-sized direct detection experiments.
}
\label{fig:mamu}
\end{figure}
\begin{figure}[!t]
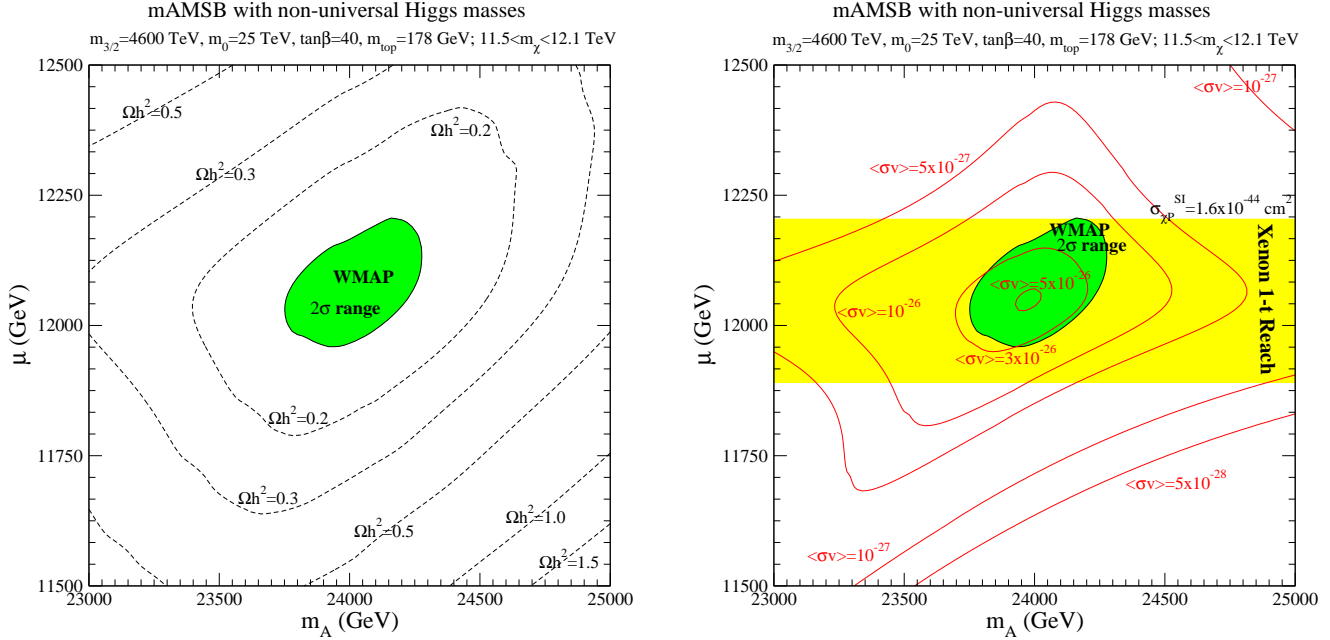

\begin{center}
\hspace*{-0.7cm}\mbox{\epsfig{file=amsb_oh2.eps,height=8.5cm}\quad\quad\epsfig{file=amsb_sv.eps,height=8.5cm}}
\end{center}
\caption{\it\small  As in Fig.~\protect{\ref{fig:mamu}}, but for neutralinos in the mAMSB model with general soft supersymmetry breaking Higgs masses.
}
\label{fig:amsb}
\end{figure}
As sample illustrative cases, we considered, in Fig.~\ref{fig:mamu} and \ref{fig:amsb} two NUHM parameter space slices along the $(m_A,\mu)$ plane, fixing
\begin{equation}
m_0=25 \ {\rm TeV},\ M_{1/2}=11 \ {\rm TeV},\ A_0=0,\ \tan\beta=40,\  {\rm sign}(\mu)>0
\end{equation}
for the NUHM mSUGRA case (Fig.~\ref{fig:mamu}), and 
\begin{equation}
\tan\beta=40,\  m_{3/2}=4600 \ {\rm TeV},\ m_0=25 \ {\rm TeV},\ {\rm sign}(\mu)>0.
\end{equation}
for the NUHM mAMSB case. The neutralino masses range, in Fig.~\ref{fig:mamu}, from 4.5 to 5.2 TeV, and in Fig.~\ref{fig:amsb} between 11.5 and 12.1 TeV, well above the quoted maximal mSUGRA and mAMSB neutralino masses. The left parts of the Figures show the contours at constant relic abundance, and feature a shaded green area corresponding to the DM abundance range of Eq.~(\ref{eq:wmap}). The neutralino mass can be read out of the plots as the minimum between $M_1\simeq5.2$ TeV (resp. $M_2\simeq12.1$ TeV) and $\mu$ (on the $y$-axis). In both cases, we see that the lowest relic abundance corresponds to the resonant $m_\chi\simeq m_A/2$ condition, and to the maximal higgsino-gaugino mixing region, {\em viz.} $M_{1,2}\simeq\mu$. The right panels indicate the contours of constant $\langle\sigma v\rangle$, and the projected reach of future, Stage-3 ton-sized direct detection experiments, such as Xenon 1-t \cite{Aprile:2004ey} (yellow shaded regions). In both cases, the maximal pair annihilation cross section region and the WMAP relic abundance areas largely fall within the reach of direct detection experiments, despite the corresponding neutralino masses being huge, respectively around 5 and 12 TeV. Though remarkable, this is not surprising, since the scalar neutralino scattering cross section off protons is mediated, besides squark exchanges, by $t$-channel $CP$-even Higgses exchanges. In the region of large higgsino-gaugino mixing these latter channels have enhanced couplings, analogously to what happens for the neutralino pair annihilation resonant $CP$-odd Higgs exchange. This points to the conclusion that, in these models, {\em although multi-TeV neutralinos will be (kinematically) beyond the reach of future accelerators, a crossing symmetry between neutralino pair annihilation processes and neutralino scattering off matter, enforced by the requirement of a sufficiently low relic abundance, will imply detectability at future direct detection experiments}. Further, comparing the right panels of Fig.~\ref{fig:mamu} and \ref{fig:amsb} with Fig.~\ref{fig:mx_sv_all}, we notice that the range of masses and annihilation cross sections nicely falls within the preferred HESS range for the NUHM mAMSB model, and (extrapolating $\langle\sigma v\rangle\approx {\rm const}\times m_\chi^{-2}$) within the Cangaroo-II range for the NUHM mSUGRA model. Concerning the spectral functions resulting from NUHM mAMSB models at large neutralino masses, we notice that the sufficiently large branching ratios into tau pairs, particularly at large $\tan\beta$, yield a rather hard spectrum at $E_\gamma\rightarrow m_\chi$, nicely compatible with the 2004 HESS data.

\section{The largest annihilation cross sections}\label{sec:ann}

{\em What is the largest possible neutralino pair annihilation cross section, at a given neutralino mass, in the minimal supersymmetric extension of the standard model?} 

\begin{figure}[!t]
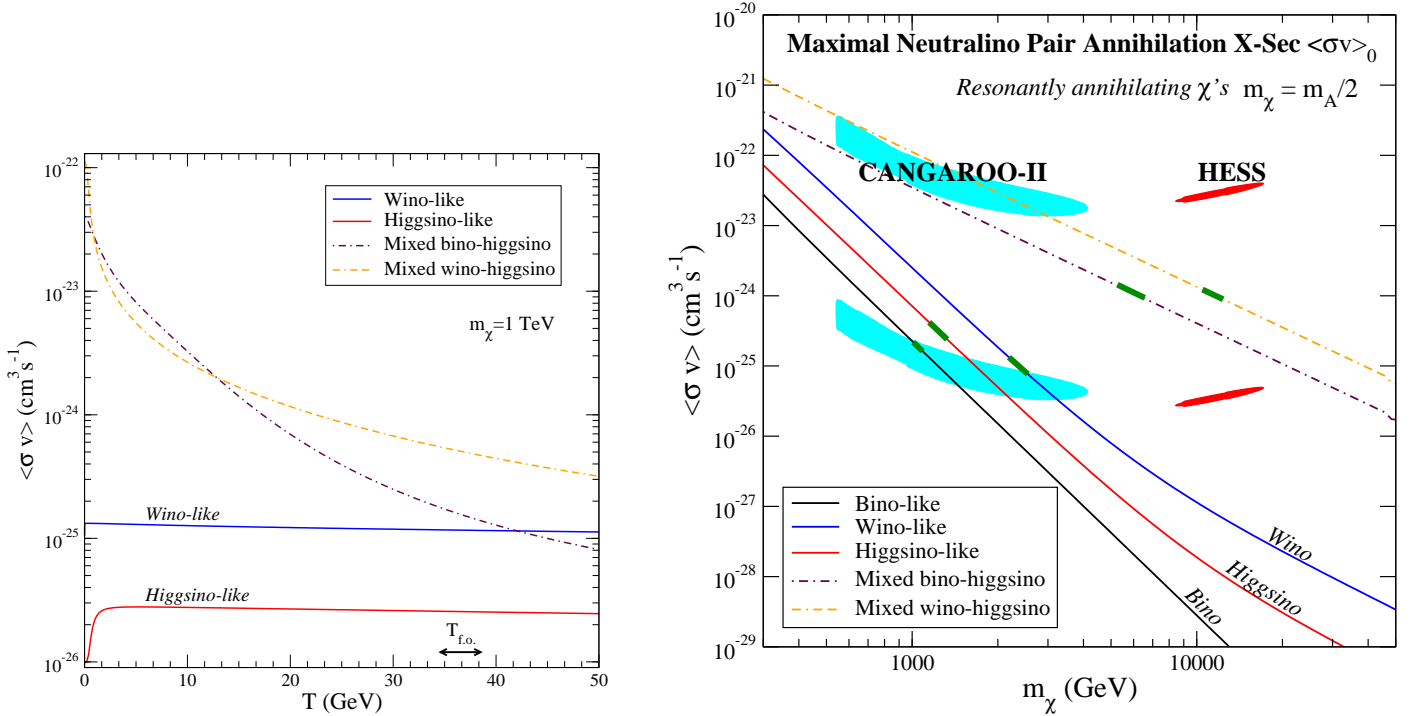

\begin{center}
\hspace*{-0.7cm}\mbox{\epsfig{file=sv_rescoan_new.eps,height=7.5cm}\quad\quad\epsfig{file=mx_sv_res_new.eps,height=9.5cm}}
\end{center}
\caption{\it\small  (Left): The thermally averaged effective cross section $\langle\sigma_{\rm eff} v\rangle(T)$, at a neutralino mass of 1 TeV, for four models (respectively a mAMSB wino-like neutralino, a mSUGRA neutralino lying in the HB/FP region, and two resonantly annihilating models at $m_1=\mu$ (mixed bino-higgsino) and $m_2=\mu$, such that $m_\chi=m_A/2$.).
(Right): cross sections of resonantly annihilating neutralinos in the $(m_\chi,\langle\sigma v\rangle)$ plane. The green strips correspond to models giving a thermal neutralino relic abundance within the 2-$\sigma$ WMAP range.
The light blue and red shaded areas refer to the 90\% C.L. regions for the Cangaroo-II data and for the 2003+2004 HESS data, for a NFW profile (two upper shaded areas) and for the adiabatically contracted N03 profile
(two lower shaded areas). We assume here that neutralinos pair annihilate with BR($b\bar b$)=70\% and BR($\tau^+\tau^-$)=30\%.}
\label{fig:rescoan}
\end{figure}
Despite the MSSM parameter space being so large, one can make general statements regarding the supersymmetric configurations featuring the largest possible values of $\langle\sigma v\rangle$. For instance, given a certain particle spectrum, since binos do not couple to gauge bosons, and feature a suppressed $s$-wave amplitude, the neutralino LSP composition of the models giving the largest possible annihilation cross sections, banning special cases, as resonant annihilations, will be mostly wino or higgsino-like, or a mixture of the two. We saw above that the sfermion spectrum is not critical for the wino/higgsino-like pair annihilation cross section: the chargino or neutralino $t$-channel exchange amplitudes always dominate over sfermion exchange ones. The remaining possibility therefore pertains to the Higgs sector, and to the occurrence of $s$-channel resonances (see Ref.~\cite{Gondolo:1990dk} for an accurate account of the treatment of resonant annihilation channels). The only MSSM particle giving a non-vanishing $s$-wave amplitude is the $CP$-odd $A$ Higgs boson \cite{Jungman:1995df}. Needless to say, the maximal resonant effect will be obtained at $m_\chi=m_A/2$, and, as discussed above, at maximal higgsino-gaugino mixing. In order to compare the pair annihilation cross section of resonantly annihilating, maximally gaugino-higgsino mixed neutralinos with the standard effective cross section of winos and higgsinos, we show in Fig.~\ref{fig:rescoan}, left panel, the thermally averaged $\langle\sigma_{\rm eff} v\rangle(T)$ (see Ref.~\cite{Edsjo:1997bg} for the definition of $\sigma_{\rm eff}$, {\em i.e.} the effective cross section including properly weighed coannihilation effects), at a neutralino mass of 1 TeV, for four models (respectively a mAMSB wino-like neutralino, a mSUGRA neutralino lying in the HB/FP region, and two resonantly annihilating models at $m_1=\mu$ (mixed bino-higgsino) and $m_2=\mu$, with $m_\chi=m_A/2$). Evidently, models with an open resonant annihilation mode give the largest cross sections at $T=0$.

In order to have an analytical insight, let us consider $\langle\sigma v\rangle$ in the limit of zero relative velocity, at $m_\chi=m_A/2$
\begin{equation}\label{eq:sigvmax}
\langle\sigma v\rangle(T=0)=\frac{g^2_{\chi\chi A}}{8\pi\Gamma_A^2}\ \sum_f c_f \beta_f |g_{ffA}|^2,\qquad \beta_f\equiv\sqrt{1-\frac{m_f^2}{m_\chi^2}},
\end{equation}
where $f$ stands for the final state fermions, $c_f$ is a color factor equal to 3 in case of quarks and to 1 in the case of leptons, $g_{\chi\chi A}$ and $g_{ffA}$ respectively indicate the neutralino pair and fermion-antifermion couplings to the $CP$-odd Higgs boson $A$, and $\Gamma_A$ is the $A$ total width. Disregarding QCD corrections, we can write
\begin{equation}
\Gamma_A\simeq\frac{m_\chi}{4\pi}\ \sum_f c_f\beta^\prime_f|g_{ffA}|^2,\qquad \beta_f^\prime\equiv\sqrt{1-\left(\frac{4m_f}{m_\chi}\right)^2},
\end{equation}
and
\begin{equation}
g_{f_df_dA}=\frac{g_2m_f}{2m_W}\tan\beta,\qquad g_{f_uf_uA}=\frac{g_2m_f}{2m_W}\frac{1}{\tan\beta}
\end{equation}
\begin{equation}\label{eq:chichia}
g_{\chi\chi A}=\left(g_2Z_{12}-g_1Z_{11}\right)\left(Z_{14}\cos\beta-Z_{13}\sin\beta\right),
\end{equation}
using the following convention for the lightest neutralino composition:
\begin{equation}
\chi=Z_{11}\widetilde{B}+Z_{12}\widetilde{W}^3+Z_{13}\widetilde{H}^0_d+Z_{14}\widetilde{H}^0_u.
\end{equation}
Neglecting the two lighter fermion generations, and assuming a heavy enough neutralino (namely, $m_\chi\gg m_{t}$), in order to maximize Eq.~(\ref{eq:sigvmax}), in view of the mild variations with $\tan\beta$ of Eq~(\ref{eq:chichia}), we need to find the maximum, as a function of $\tan\beta$, of
\begin{equation}
\left((3m_b^2+m_\tau^2)\tan^2\beta+3m_t^2/\tan^2\beta\right)^{-1}
\end{equation}
which is found to occur at 
\begin{equation}
\tan\beta\approx\sqrt{m_t/m_b}\approx 6\div7.
\end{equation}
Plugging the value of the minimum of $\Gamma_A$ into Eq.~(\ref{eq:sigvmax}), we give the following estimate,
\begin{equation}\label{eq:svaest}
\langle\sigma v\rangle(T=0)\approx\frac{2\pi}{m_\chi^2}\frac{2m_W^2}{3m_t\cdot m_b}\left(\frac{g_{1,2}}{g_2}\right)^2\left( Z_h\cdot Z_{g_{1,2}} \right),
\end{equation}
the subscripts 1 and 2 referring to the case $m_1<m_2$ and $m_1>m_2$ respectively ({\em i.e.} to a mixed bino or wino LSP), and $Z_h=Z_{13}^2$, $Z_{g_{1,2}}=Z_{11,12}^2$. Inserting the numerical values, we obtain
\begin{equation}\label{eq:svnest}
\langle\sigma v\rangle(T=0)\approx\ \frac{\mathcal{O}(1\div10)}{m_\chi^2}\ {\rm GeV}^{-2}\ \sim\ 10^{-23}\div10^{-22}\ \left(\frac{m_\chi}{1\ {\rm TeV}}\right)^{-2}\ \ {\rm cm}^3{\rm s}^{-1}.
\end{equation}
Eq.~(\ref{eq:svaest}) shows that the maximal annihilation cross section occurs for maximally mixed (last factor) wino-higgsino neutralinos (since $g_2>g_1$). The maximal bino-higgsino annihilation cross section will be suppressed, according to Eq.~(\ref{eq:svaest}), by a factor $(g_1/g_2)^2=\tan^2\theta_W\approx0.3$.

The above predictions are nicely confirmed by our numerical results, showed in the right panel of Fig.~\ref{fig:rescoan} on the $(m_\chi,\langle\sigma v\rangle)$ plane. We picked three ``{\em pure}'' and two ``{\em maximally mixed}'' MSSM neutralinos, setting the (largely unimportant) common sfermion mass to $m_{\widetilde S},\ =10\times m_\chi$, $m_A=2\times m_\chi$, and the following relations for the gaugino masses and for the $\mu$ term, at the weak scale,
\begin{eqnarray}
\nonumber && \mu=10\times m_1,\ m_2=2\times m_1,\ m_3=6\times m_1 \qquad ({\rm Bino-like,\ mSUGRA\ relations})\\
\nonumber && m_1=10\times \mu,\ m_2=2\times m_1,\ m_3=6\times m_1 \qquad ({\rm Higgsino-like,\ mSUGRA\ relations})\\
\nonumber && \mu=10\times m_2,\ m_1=3.3\times m_2,\ m_3=-8\times m_2 \qquad ({\rm Wino-like,\ mAMSB\ relations})\\
\nonumber && \mu=m_1,\ m_2=2\times m_1,\ m_3=6\times m_1 \qquad ({\rm Mixed\ Bino-Higgsino,\ mSUGRA\ relations})\\
\nonumber && \mu=m_2,\ m_1=3.3\times m_2,\ m_3=-8\times m_2 \qquad ({\rm Mixed\ Wino-Higgsino,\ mAMSB\ relations})
\end{eqnarray}
For all models, the value of $\tan\beta$ was chosen, at each $m_\chi$ in order to numerically maximize $\langle\sigma v\rangle$, and found to be always close to our estimate above, $\tan\beta\approx 7$, except for $m_\chi\lesssim4\times m_t$, where even lower values of $\tan\beta$ gave larger cross sections. In the Figure, we indicate with a thick green line the range of masses giving a thermal relic abundance within the WMAP range of Eq.~(\ref{eq:wmap}). As a side remark, notice how badly the naive relation quoted in Eq.~(\ref{eq:naiverelation}) fails for resonantly annihilating models. The light blue and red shaded areas correspond to the 90\% C.L. fit countours for the Cangaroo-II data and for the 2003+2004 HESS data (computed for a final state with  BR($b\bar b$)=70\% and BR($\tau^+\tau^-$)=30\%, indicative of the final state pattern of resonantly annihilating neutralinos). We notice that {\em the HESS data can have a supersymmetric interpretation with a model producing the correct thermal neutralino relic abundance, and with a boost factor of ${\mathcal{O}(10)}$}.

Assuming a GUT unification relation for the gaugino masses, a mixed wino-higgsino state cannot be realized. Henceforth we conclude this Section quoting the two relevant $\langle\sigma v\rangle$ theoretical upper limits obtained:
\begin{eqnarray}
&& \langle\sigma v\rangle^{\rm \small MAX}\ \simeq\ 10^{-22}\ \left(\frac{m_\chi}{1\ {\rm TeV}}\right)^{-2}\ \ {\rm cm}^3{\rm s}^{-1}\qquad\quad\ \ {\bf no\ gaugino \ mass\ unification}\label{eq:thubwh}\\
&& \langle\sigma v\rangle^{\rm \small MAX}\ \simeq\ 3\times 10^{-23}\ \left(\frac{m_\chi}{1\ {\rm TeV}}\right)^{-2}\ \ {\rm cm}^3{\rm s}^{-1}\qquad {\bf with\ gaugino \ mass\ unification}\label{eq:thubbh}
\end{eqnarray}
We stress that the results quoted above apply to neutralino masses $m_\chi\gtrsim 100$ GeV, since all the models we considered feature a quasi-degenerate chargino, whose mass is bounded to be larger than approximately 103 GeV \cite{Eidelman:2004wy}.

\subsection{The role of non-perturbative EW effects}\label{sec:npew}

\begin{figure}[!t]
\begin{center}
\hspace*{-1cm}\epsfig{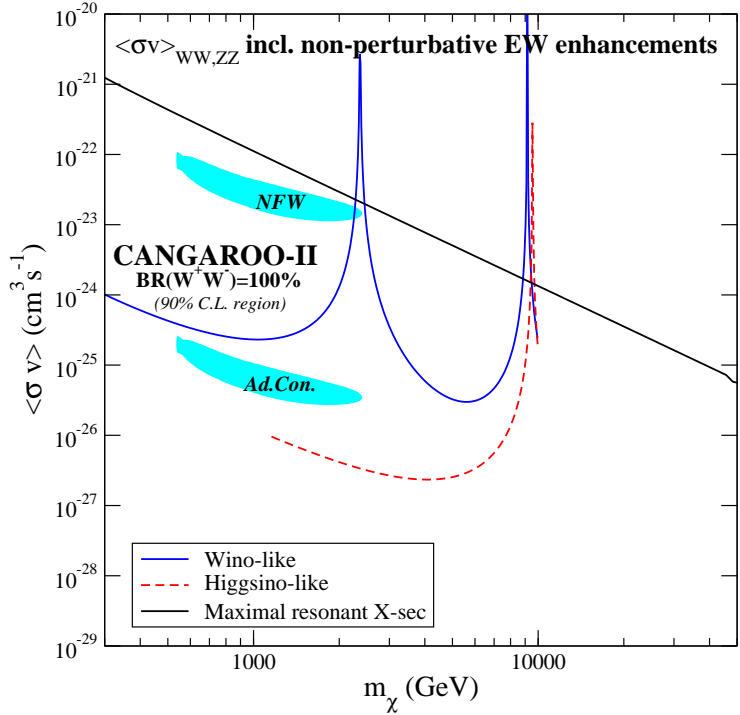}
\end{center}
\caption{\it\small  The $\langle\sigma v\rangle$ contributing to the continuum gamma-rays yield from neutralino pair-annihilations including non-perturbative electro-weak effects \cite{Hisano:2004ds}, for a wino-like (solid blue line) and for a higgsino-like (dashed red line) lightest neutralino. The chargino-neutralino mass splitting was computed including loop effects from custodial $SU(2)$-breaking in the gauge boson sector, {\em i.e.} in the limit of heavy sfermions \cite{Cheng:1998hc}. We also indicate the 90\% C.L. regions for the Cangaroo-II data set fit, for neutralinos annihilating into gauge bosons. We recall that the pure gauge bosons final state is excluded by the 2004 HESS data (see Sec.~\ref{sec:act}).
}
\label{fig:npew}
\end{figure}

Recently, it has been noticed that $SU(2)_L$ non-singlet neutralinos ({\em i.e.} wino or higgsino-like states), featuring a quasi-degenerate chargino partner, with mass in the TeV range or above, undergo, at small velocities, large enhancements in their pair annihilation cross section \cite{Hisano:2002fk,Hisano:2004ds,Boudjema:2005hb}. This is due to non-perturbative effects, leading to the formation of multiple neutralinos-charginos ``bound states'' which can resonantly contribute to the annihilation cross section in the non-relativistic limit \cite{Hisano:2004ds}. In Ref.~\cite{Hisano:2002fk,Hisano:2004ds} a non-relativistic effective action for $SU(2)_L$ doublets (Higgsinos) and triplets (Winos) was derived, leading to a numerical evaluation of the pair annihilation cross section of neutralinos into final state gauge bosons $VV^\prime$. This non-perturbative contribution was never taken into account before, and a convenient numerical fitting formula for $\langle\sigma v\rangle_{VV^\prime}$, as a function of the neutralino mass and of its mass splitting $\delta m$ with its chargino partner, was provided as well \cite{Hisano:2004ds}. Evidently, non-perturbative electroweak effects go beyond the treatment outlined above, and must therefore be analyzed to assess if and where the theoretical upper bounds on $\langle\sigma v\rangle$ quoted in Eq.~(\ref{eq:thubwh}),(\ref{eq:thubbh}) are violated.

In the present context, we sum over the gauge boson final states contributing to the gamma rays continuum, namely $W^+W^-$ and $ZZ$. We assume, for definiteness, and to get a model-independent result, that the dominant loop corrections entering $\delta m$ come dominantly from the gauge bosons loop contributions \cite{Cheng:1998hc,Giudice:1995qk}, which only depend on the neutralino mass. We show our results in Fig.~\ref{fig:npew}. The thick solid blue and the red dashed lines reproduce the numerical fitting formula of Eq.~(61) in {\tt hep-ph/0412403v2} within their range of validity. The black line indicates the (``perturbative'') upper limit quoted above in Eq.~(\ref{eq:thubwh}). Evidently, when $m_\chi$ hits those special values (depending on the chargino-neutralino splitting) corresponding to ``zero energy'' resonances, $\langle\sigma v\rangle$ locally exceeds the perturbative result of Eq.~(\ref{eq:thubwh}). However, this happens in very limited neutralino mass ranges. Further, it would be very difficult to reliably predict the value of $\langle\sigma v\rangle(m_\chi)$, which is a very steeply varying function, close to the resonances. The non-perturbative electro-weak resonance scenario appears in any case unfit to explain the HESS data. In fact, as pointed out in Sec.~\ref{sec:act}, the resulting pure gauge bosons final state spectral function for the gamma rays continuum (even including the internal bremsstrahlung effects of Ref.~\cite{Bergstrom:2005ss}) is statistically excluded by the analysis of the 2004 HESS data. 

\section{The largest masses: non-perturbative QCD effects and gluino coannihilations}\label{sec:gluino}

In the previous Section we found that neutralinos with the largest possible pair annihilation cross section produce a thermal relic abundance compatible with the WMAP upper bound quoted in Eq.~(\ref{eq:wmap}) for masses $m_\chi\lesssim 12.5$ TeV. It is well known, however, that when a particle exists that can pair annihilate (``{\em co-annihilate}'') with the lightest neutralino and that is close in mass with it, so that their freeze-out processes overlap and interfere in the early Universe, the final thermal relic abundance of the neutralino can be largely affected \cite{GriestSeckel}. We therefore expect that the largest neutralino mass compatible with a thermal production of DM can be found in models where coannihilation processes are active. The $s$-wave unitarity limit on a thermally produced WIMP mass derived in Ref.~\cite{Griest:1989wd}, suitably rescaled with the result of Ref.~\cite{Spergel:2003cb} quoted in Eq.~(\ref{eq:wmap}) leads to a putative upper bound of $m_\chi\lesssim120$ TeV. However, the $s$-wave unitarity argument does not always apply (a counter example is given for instance by resonantly annihilating WIMPs, as those considered in the previous Section). We wish to assess here which is the maximal theoretically allowed neutralino mass in the general MSSM, and whether or not the unitarity bound quoted above applies.

As a rule of thumb, if the neutralino pair annihilation cross section is smaller than the coannihilating partner pair annihilation cross section times the ratio of the number of internal degrees of freedom of the neutralino over that of the coannihilating partner\footnote{For a more precise condition, including the co-annihilation cross sections contribution, see Ref.~\cite{Edsjo:2003us}} the neutralino relic abundance in presence of coannihilations will be {\em reduced}. Again as a rule of thumb, the Maxwell-Boltzmann equilibrium distribution fixes the relative ``weight'' of the coannihilation contribution to the effective cross section determining the neutralino relic abundance \cite{GriestSeckel} as scaling $\propto\exp(-\Delta m/T_{\rm f.o.})$, where $\Delta m$ is the mass splitting between the neutralino and the coannihilating partner, and $T_{\rm f.o.}$ is the neutralino freeze-out temperature.

Many {\em minimal} supersymmetry breaking models include the possibility of coannihilation processes, for instance the mSUGRA and the mAMSB scenarios discussed above. It goes without saying that the largest the pair annihilation cross section of the coannihilating partner, the largest the maximal possible coannihilation effects. In Ref.~\cite{Profumo:2004wk} a model-independent analysis showed that, not surprisingly, the ``strongest'' coannihilating partners in the MSSM are strongly interacting particles (SIPs), the squarks and the gluino. Stop coannihilations were considered previously in Ref.~\cite{Boehm:1999bj,stop,Edsjo:2003us}. The analysis of Ref.~\cite{Profumo:2004wk} was based on the {\tt micrOMEGAs} code for the relic density computation \cite{micromegas}, which includes the perturbative cross section only for SIPs. Already at the perturbative level, gluinos were found to be by far more efficiently coannihilating than squarks. It was also noticed, however, that special MSSM realizations, including the occurrence of {\em multiple} squark coannihilations or of resonantly annihilating coannihilation partners, could yield net effects on the neutralino relic abundance even beyond that of the (perturbative) gluino coannihilations. To quantitatively assess this possibility, we carried out an extensive scan of the general MSSM parameter space, along the lines of Ref.~\cite{Profumo:2004at} (where the reader is directed for further details). We find that the above mentioned special, extremely fine-tuned models where peculiar squark coannihilations occur can suppress the neutralino relic abundance to the appropriate DM density level for neutralinos as heavy as 20$\div$25 TeV.

The inclusion of non-perturbative strong interactions effects in SIPs pair annihilations clearly plays a crucial role for an accurate neutralino relic abundance computation, and henceforth to nail down the maximal neutralino mass compatible with a thermal relic abundance in the CDM range. In Ref.~\cite{Boehm:1999bj} it was claimed that a full higher order QCD calculation in the stop (co-)annihilation cross sections can amount to corrections around a factor 2 or so with respect to the perturbative result. Strong interactions effects are, on the other hand, expected to be much more relevant in the case of gluino pair annihilations. With this in mind, we consider below the case of gluino (co-)annihilations, motivated by the available results of the extended and accurate analysis of the non-perturbative QCD effects in the gluino pair annihilation cross sections carried out in Ref.~\cite{Baer:1998pg}.

Supersymmetric scenarios where gluinos can coannihilate with neutralinos include the widely discussed class of models featuring non-universal gaugino masses \cite{nugm}. A string-inspired model which favors a scenario where the gluino can be either the LSP or a (possibly coannihilating) next-to-LSP is the O-II string model in the limit where supersymmetry breaking is dominated by the universal ``size'' modulus \cite{oII}. A recently proposed framework where gluinos can be coannihilating partners of a neutralino LSP is that of Split-Supersymmetry, where the superpartners scalar sector is assumed to be ``split'' from the fermionic sector \cite{Arkani-Hamed:2004fb,splitsusy}. Other specific theoretically motivated models were discussed in Ref.~\cite{Baer:1998pg,Profumo:2004at}.

The publicly available numerical package for the computation of the neutralino relic density {\tt DarkSUSY} \cite{Gondolo:2004sc} does not include gluino coannihilations, while, as mentioned above, the {\tt micrOMEGAs} package \cite{micromegas} does not go beyond the leading perturbative cross sections for gluino (co-)annihilations. We therefore developed an independent numerical code to account for gluino coannihilations, including various non-perturbative gluino pair-annihilation scenarios, and we interfaced it with the $\langle\sigma_{\rm eff} v\rangle(T)$ computation of the {\tt DarkSUSY} package to compute the neutralino relic abundance, hence automatically taking into account all the effects of other coannihilating partners, of resonances and of thresholds. We do not include higher order QCD effects in the gluino-neutralino cross section, since we do not expect them to be quantitatively relevant. Further, as already observed in Ref.~\cite{Profumo:2004at}, even considering a fully perturbative $\widetilde g\widetilde g$ annihilation cross section, the gluino-neutralino coannihilation term is always sub-dominant in the $\langle\sigma_{\rm eff} v\rangle(T)$ computation (it was found there to give at most a per-cent contribution). We stress here, however, that gluino-neutralino conversion processes are essential to keep gluinos in thermal equilibrium with neutralinos through scattering off relativistic quarks during freeze-out. Since those processes are mediated by squark exchanges, this means that squarks cannot be exceedingly heavy, not to effectively decouple the gluino and neutralino freeze-out. 

The dynamics of gluino pair annihilations, and the related issue of the relic abundance of a gluino LSP, was widely discussed in the literature \cite{othergluino,Baer:1998pg}. The perturbative gluino pair annihilation cross section as a function of the center-of-mass energy $s$ reads, in the notation of Ref.~\cite{Baer:1998pg}
\begin{equation}\label{eq:sigmaglue}
\sigma_P=\sigma(\gluino\gluino\rightarrow g g)+\sum_q\sigma(\gluino\gluino\rightarrow q\overline{q})
\end{equation}
where
\begin{eqnarray}
\nonumber && \nonumber \sigma(\gluino\gluino\rightarrow g g)= \frac{3\pi\alpha_s^2}{16\beta^2s}\left(\log\frac{1+\beta}{1-\beta}(21-6\beta^2-3\beta^4)-33\beta+17\beta^3\right), \qquad \beta=\sqrt{1-4m_\gluino^2/s}\\
&& \sigma(\gluino\gluino\rightarrow q\overline{q}) = \frac{\pi\alpha_s^2\overline{\beta}}{16\beta s}(3-\beta^2)(3-\overline{\beta}^2), \qquad \overline{\beta}=\sqrt{1-4m_q^2/s}.
\end{eqnarray}
In Ref.~\cite{Baer:1998pg} it has been claimed that non-perturbative effects, relevant when $\sqrt{s}\sim 2m_\gluino$, are expected to range between two extreme scenarios. In the first scenario, one only considers the effects of multiple gluon exchanges, which, neglecting the logarithmic enhancements due to soft radiation, can be parameterized by the (exponentiated form of the) the Sommerfeld enhancement factor, 
\begin{equation}
E=\frac{C\pi\alpha_s(Q)}{\beta}\left(1-\exp\left(-\frac{C\pi\alpha_s(Q)}{\beta}\right)\right),\qquad C_{gg}=1/2,\ C_{q\overline{q}}=3/2
\end{equation}
where $\alpha_s(Q)$ is evaluated at the scale of the typical momentum transfer of the exchanged soft gluons, $Q\sim\beta m_\gluino$, and the $C_{gg}$ and $C_{q\overline{q}}$ respectively apply to the gluon pair final state cross section and to the quark-antiquark pair.

In the opposite, extreme non-perturbative scenario, gluinos undergo, at small $\beta$'s, a transition into color singlet bound states. The non-perturbative pair annihilation of those states into $\pi$'s is usually assumed to be 
\begin{equation}\label{eq:sigmagluenp}
\sigma_{NP}=A\beta^{-1}/m_\pi^2,\qquad A=\mathcal{O}(1).
\end{equation} 
The location of the transition is highly uncertain; in the most extreme scenario considered in Ref.~\cite{Baer:1998pg}, the transition was taken to occur when the total $\gluino\gluino$ kinetic energy in the center of mass frame fell below a given limit $L$, with $L\sim0.2\div1$ GeV. Below the gluino-bound state transition, the scattering cross section was assumed to follow the perturbative result in Eq.~(\ref{eq:sigmaglue}), with either an abrupt or a smooth transition to the non-perturbative annihilation cross section (\ref{eq:sigmagluenp}). The sudden transition option was found, in Ref.~\cite{Baer:1998pg}, to give a smaller final gluino relic abundance, and hence we expect, with this choice, the maximal possible gluino coannihilation effects.

\begin{figure}[!t]
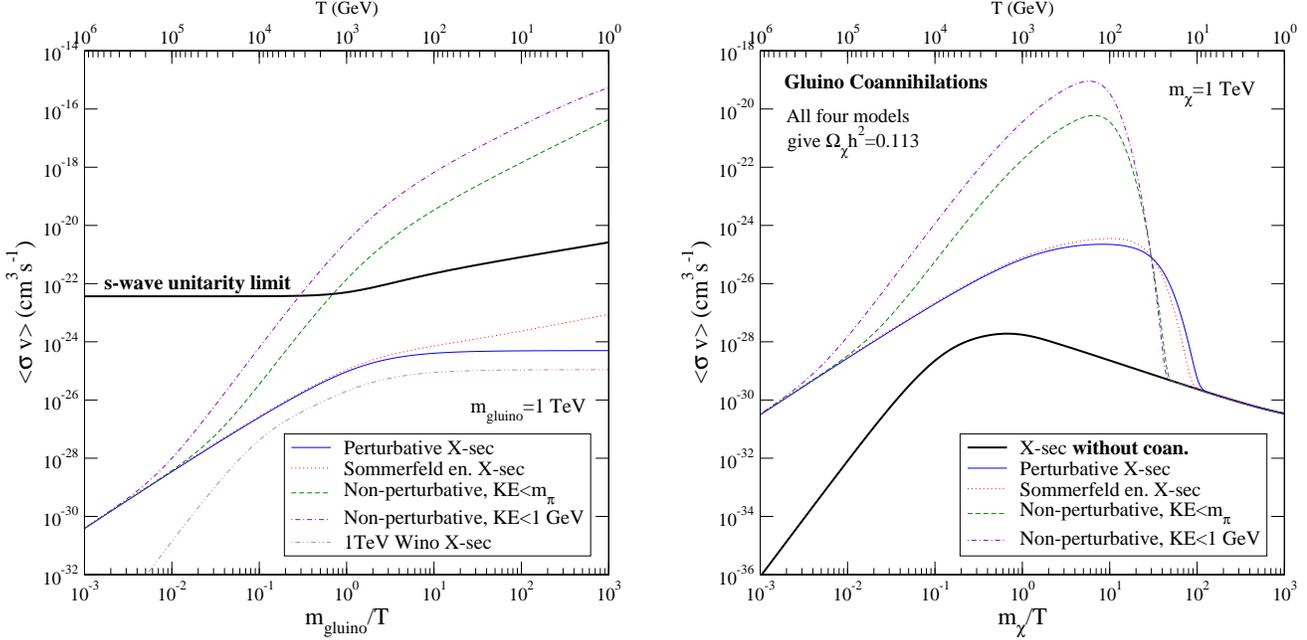

\begin{center}
\hspace*{-0.7cm}\mbox{\epsfig{file=sv_gluino.eps,height=8.5cm}\quad\quad\epsfig{file=sv_gluinocoan.eps,height=8.5cm}}
\end{center}
\caption{\it\small  (Left): the thermally averaged gluino pair annihilation cross section as a function of $m_{\gluino}/T$ (lower x axis) and of $T$ (upper $x$-axis, respectively), for a gluino mass $m_{\gluino}=1$ TeV, assuming a fully perturbative $\sigma_P$, a Sommerfeld-enhanced $E\times\sigma_P$, and a fully non-perturbative annihilation cross section. We also include the $s$-wave unitarity limit \cite{Griest:1989wd} (solid black line) and the cross section of a 1 TeV purely wino-like neutralino (brown dot-dashed line). (Right): The {\em effective} thermally averaged neutralino pair annihilation cross section $\langle\sigma v\rangle$ ({\em i.e.} including co-annihilations) as a function of $m_{\chi}/T$ (lower x axis) and of $T$ (upper $x$-axis), for a mSUGRA bino-like neutralino with a 1 TeV mass ($\tan\beta=30$, $m_0=5M_{1/2},\ \mu>0,\ A_0=0$), without gluino coannihilations (solid black line) and including gluino coannihilations. In the latter case we show the lines corresponding to the four gluino annihilation cross sections scenarios of the left panel, for gluino-neutralino mass splittings such that the resulting final neutralino relic abundance corresponds to the central non-baryonic Dark Matter abundance deduced from the WMAP data \cite{Spergel:2003cb}, $\Omega_\chi h^2=0.113$ .
}
\label{fig:gluinocoan}
\end{figure}
We show the gluino pair annihilation cross sections we consider here in Fig.~\ref{fig:gluinocoan}, left panel. There, we plot $\langle\sigma v\rangle$ as a function of $m/T$ and of $T$ (lower and upper $x$-axis), for a gluino mass of 1 TeV. We picked two options for the non-perturbative scenario, featuring a transition at a kinetic energy of (1) 1 GeV and (2) equal to the $\pi$ mass. We also plot the $s$-wave unitarity limit and the annihilation cross section of a 1 TeV mAMSB wino, for comparison. The extreme non-perturbative scenario above violates the $s$-wave unitarity limit on the pair annihilation cross section of a WIMP, as already pointed out in Ref.~\cite{Baer:1998pg}, as a consequence of coherent contributions from multiple partial waves. It can be noted that at large temperatures all scenarios converge to the perturbative cross section, but at low temperatures the differences in the various pair annihilation options can be substantial.

\begin{figure}[!t]
\begin{center}
\epsfig{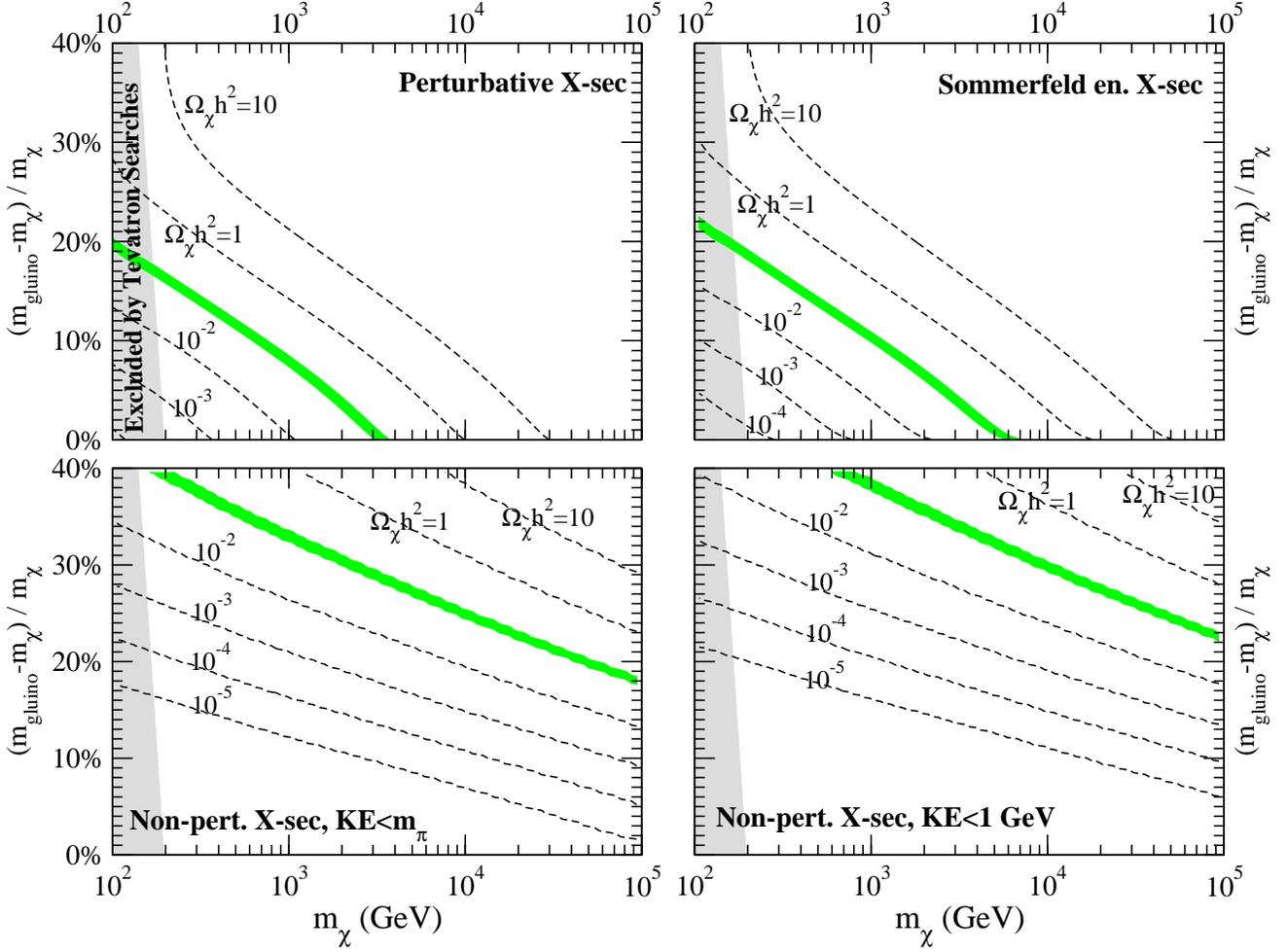}
\end{center}
\caption{\it\small  Iso-level curves corresponding to given neutralino relic abundance values, in the plane defined by the neutralino mass ($x$-axis) and by the relative gluino-neutralino mass splitting ($y$-axis), for four different gluino pair annihilation scenarios. The green shaded area corresponds to the 95\% C.L. range for $\Omega_\chi h^2$, while the area shaded in grey, in the right part of each panel, indicates the region in the plane excluded by Tevatron searches for gluinos \cite{Eidelman:2004wy}.
}
\label{fig:gluinooh2}
\end{figure}
The importance of including gluino coannihilations for the relic abundance of neutralinos can be appreciated in the right panel of Fig.~\ref{fig:gluinocoan}. We consider, there, a 1 TeV bino-like neutralino taken from the mSUGRA parameter space, at $m_0=5M_{1/2}$, $\mu>0$, $\tan\beta=30$ and $A_0=0$, and we indicate its thermally averaged cross section with a thick black solid line. The relic abundance of that neutralino without gluino coannihilations would be $\Omega_\chi h^2\simeq27$. We then add a coannihilating gluino, featuring a pair annihilation cross section following one of the four schemes of Fig.~\ref{fig:gluinocoan}, left, with a mass such that the resulting neutralino relic abundance gives the central WMAP value, $\Omega_\chi h^2\simeq0.113$ (this amounts to considering gluino-neutralino mass splittings ranging from 7\% in the fully perturbative case to 38\% in the extreme non-perturbative case with transition at a kinetic energy of 1 GeV). We then plot the resulting {\em effective} neutralino annihilation cross section , as defined in Ref.~\cite{Edsjo:1997bg}, {\em i.e.} the cross section including all the properly weighed (co-)annihilation contributions. We see that around the freeze-out temperature, the effective annihilation cross section including coannihilations exceeds the neutralino pair annihilation cross section by even four orders of magnitudes; also notice that in order not to exceedingly suppress the final neutralino relic abundance, in the extreme non-perturbative scenarios the mass splitting is much larger, and hence the temperature at which gluino coannihilations effectively disappear is also larger.

To assess the maximal neutralino mass compatible with the WMAP result on the DM abundance in the neutralino-gluino coannihilation scenario, we show contour levels of iso-relic abundance in the $(m_\chi,(m_\gluino-m_\chi)/m_\chi)$ plane in Fig.~\ref{fig:gluinooh2}, for the four gluino pair annihilation scenarios of Fig.~\ref{fig:gluinocoan}. The gray shaded area in the left part of each panel indicates gluino masses excluded by Tevatron searches \cite{Eidelman:2004wy}, while the green band locates the 95\% C.L. range of the CDM abundance as determined by WMAP \cite{Spergel:2003cb}. While in the most conservative non-perturbative scenario (Sommerfeld enhanced cross section, top right panel) the maximal neutralino mass is even below 10 TeV, the two extreme non-perturbative frameworks allow for neutralino masses well above 100 TeV, without invoking a terribly small mass splitting between the gluino and the neutralino mass (the mass splittings in the stau coannihilation region are always well below what we consider here).

In the extremely non-perturbative gluino coannihilation scenarios considered in the two panels at the bottom, the violation of the $s$-wave unitarity limit in gluino pair annihilations (Fig.~\ref{fig:gluinocoan}, left) is effectively ``{\em transferred}'' to the neutralino sector. As a consequence, in those scenarios, neutralinos with masses at, or even above 100 TeV produce a sufficiently low thermal relic abundance, provided the gluino has a mass within, say, 20\% of the neutralino mass. 

As a bottom line, in view of our discussion above, we conclude that {\em taking into account non perturbative effects in gluino pair annihilations, and depending on the assumed non-perturbative scenario, the largest neutralino mass in the MSSM compatible with a WMAP relic abundance ranges from around 20 TeV to well above 100 TeV}.

\section{Outlook: the largest ``supersymmetric factors'' for neutralino indirect detection}\label{sec:outlook}

We provided in Sec.~\ref{sec:ann} an analytical and numerical estimate of the theoretical upper bound on the neutralino pair annihilation cross section $\langle\sigma v\rangle$, as a function of its mass. Neutralinos featuring that $\langle\sigma v\rangle$ produce a relic abundance in the WMAP range only at very large masses, around 12 TeV. At smaller masses, those models do not thermally produce enough relic neutralinos to explain the observed DM abundance. In that case, one either assumes that neutralinos only make up for a fraction of DM, the rest being composed by some other particles, or relaxes the assumptions that lead to the usual thermal relic abundance result. In this latter approach, as already mentioned, a few loopholes have been considered so far: neutralinos can be produced non-thermally in the decay of moduli, gravitinos, Q-balls, cosmic strings etc. \cite{nonth}, or a modified behavior for the Hubble expansion factor $H$ (driving the neutralino freeze-out) might have occurred, affecting the final relic density \cite{Kamionkowski:1990ni,Profumo:2004ex}. Various examples of modified cosmologies have been proposed so far (including quintessential scenarios \cite{quint}, Brans-Dicke-Jordan cosmologies \cite{Kamionkowski:1990ni,bdj}, anisotropic cosmologies \cite{Kamionkowski:1990ni,Profumo:2004ex} etc.). 

\begin{figure}[!t]
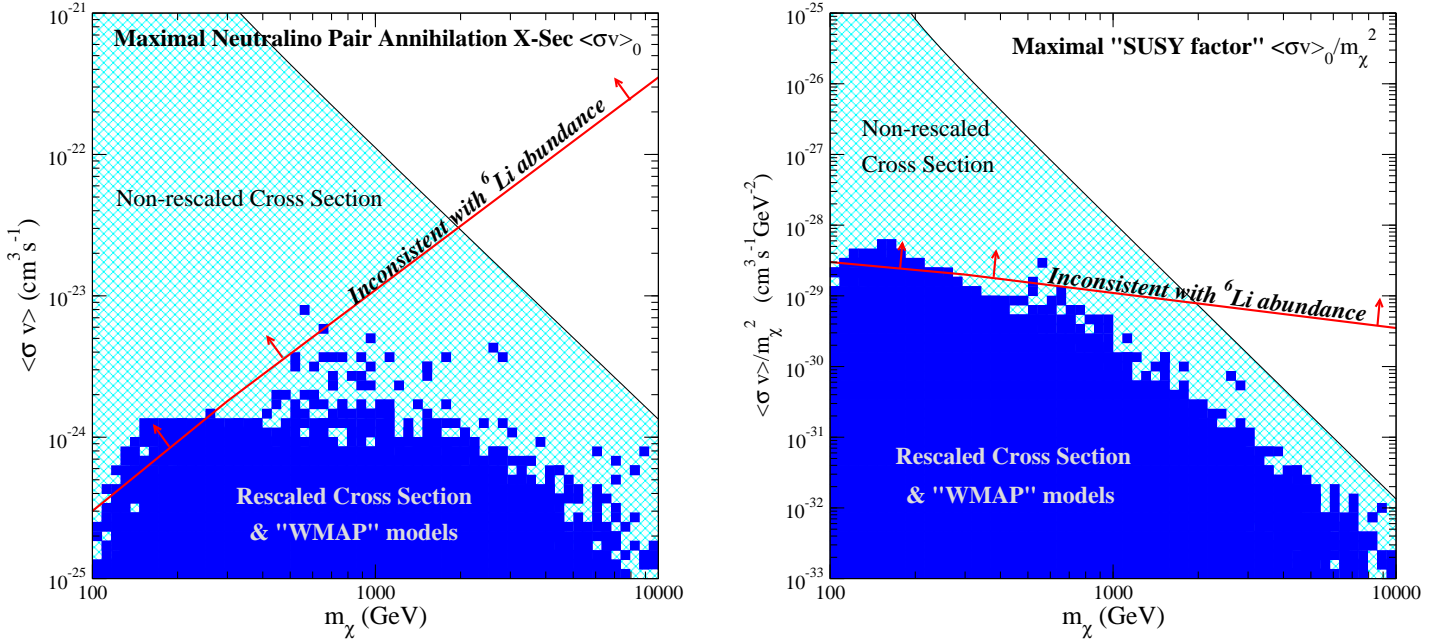

\begin{center}
\hspace*{-0.7cm}\mbox{\epsfig{file=mx_sv_max.eps,height=8.5cm}\quad\quad\epsfig{file=mx_svm2.eps,height=8.5cm}}
\end{center}
\caption{\it\small  The regions, in the $(m_\chi,\langle\sigma v\rangle)$ plane, populated by supersymmetric models with non-rescaled cross section and arbitrarily low relic abundances (light-blue hatched region) and by models with a WMAP relic abundance or with a {\em rescaled} cross section ({\em i.e.} if $\Omega_\chi h^2<\Omega_{\rm DM}^{\rm min} h^2$ then $\langle\sigma v\rangle\rightarrow \langle\sigma v\rangle\times (\Omega_\chi/\Omega_{\rm DM}^{\rm min})^2$), indicated by dark blue squares. The latter were obtained through a random scan over the general, flavor diagonal and CP-conserving, MSSM parameter space. The region lying above the red line produces an excess of ${}^6$Li in the early Universe, if all DM is composed by neutralinos. (Right): same as for the left panel, but for the ``supersymmetric factor''  $\langle\sigma v\rangle/m^2_\chi$.
}
\label{fig:maxsv}
\end{figure}
In the most conservative approach, without assuming non-thermal production or cosmological relic density enhancement processes, any signal from neutralino pair annihilations will be proportional to the quantity
\begin{equation}\label{eq:rescale}
\xi^2\langle\sigma v\rangle/m^2_\chi,\qquad \xi\equiv{\rm min}[1,(\Omega_\chi h^2)/(\Omega_{\rm CDM}h^2)_{\rm min}],\quad (\Omega_{\rm CDM}h^2)_{\rm min}=0.095
\end{equation}
the factor $\xi$ taking into account the fraction of DM composed by neutralinos. In this case, it is highly non-trivial to theoretically predict which is the maximal supersymmetric factor $\xi^2\langle\sigma v\rangle/m^2_\chi$. We thus resorted to a very large numerical scan of the MSSM, including all phenomenological constraints ({\em e.g. }from supersymmetric contributions to BE($b\rightarrow s\gamma$), BR($B_s\rightarrow\mu^+\mu^-$), precision electro-weak observables, supersymmetric particles searches, Higgs searches, etc.). The scan was again performed along the lines discussed in Ref.~\cite{Profumo:2004at}, where the reader is directed for further details. As a by-product, we also explicitly verified that no MSSM models in our scan violate the theoretical upper on $\langle\sigma v\rangle$ quoted in Eq.~(\ref{eq:thubwh}). We show our results in Fig.~\ref{fig:maxsv} for the pair annihilation cross section (left) and for the ``supersymmetric factor'' (right). We also indicate the region excluded by the neutralino-induced synthesis of ${}^6$Li in the early Universe, as computed in Ref.~\cite{Jedamzik:2004ip}. The ${}^6$Li constraint strictly applies only to the dark blue region, as long as the above mentioned non-thermal production or relic density enhancement processes are not assumed. 

Most of the WMAP models featuring a large $\langle\sigma v\rangle$ have, again, a resonant annihilation channel, with a slightly off-resonance kinematical condition on $m_A\sim2\times m_\chi$. While the lightest neutralino mass WMAP models are mostly bino-like, a non-trivial higgsino component starts at $m_\chi\sim500$ GeV, giving a significant increase in the largest possible $\langle\sigma v\rangle$ values. The largest pair annihilation cross sections and supersymmetric factors we find in the range we explored ($m_\chi\gtrsim 100$ GeV) are listed below, for the convenience of the reader,
\begin{eqnarray}
&&(\xi^2\langle\sigma v\rangle)^{\rm\sss MAX}\simeq8\times 10^{-24}\ {\rm cm}^3{\rm s}^{-1}\ \ {\rm at}\ \ m_\chi\simeq550 \ {\rm GeV}\label{eq:maxsv}\\
&&\left(\xi^2\langle\sigma v\rangle/m^2_\chi\right)^{\rm\sss MAX}\simeq6\times 10^{-29}\ {\rm cm}^3{\rm s}^{-1}{\rm GeV}^{-2}\ \ {\rm at}\ \ m_\chi\simeq150\ {\rm GeV}\label{eq:maxsf}
\end{eqnarray}
A similar analysis, extending to even smaller neutralino masses than those we consider here, was recently carried out in Ref.~\cite{Bottino:2005xy}. Although we find, as in \cite{Bottino:2005xy}, that the maxima quoted in (\ref{eq:maxsv}) and (\ref{eq:maxsf}) correspond to models with $\xi=1$, the absolute maximal value for $(\xi^2\langle\sigma v\rangle)^{\rm\sss MAX}$ we find here is more than two orders of magnitude larger than what quoted in \cite{Bottino:2005xy}, even taking into account an uncertainty factor of order unity. The reason for this discrepancy can be traced back to either the slightly different MSSM parameters over which the scans have been performed, or to numerical differences in the evaluation of the pair annihilation cross section, particularly in the proximity of resonances, where we find the largest cross sections reported here. On the other hand, Ref.~\cite{Bottino:2005xy} finds larger values for the supersymmetric factor than what we quote in (\ref{eq:maxsf}), outside the range of masses considered here, at $m_\chi\ll100$ GeV. As a side remark, we would like to point out that in presence of coannihilation processes, and particularly in view of our results presented in Sec.~\ref{sec:gluino}, no direct relations link the neutralino pair annihilation cross section with its thermal relic abundance. In this respect, it is very hard to draw a minimal, guaranteed value for either $\xi^2\langle\sigma v\rangle$ or $\xi^2\langle\sigma v\rangle/m^2_\chi$, which in presence of a ``strong'' coannihilating partner can essentially be arbitrarily low.

\begin{figure}[!t]
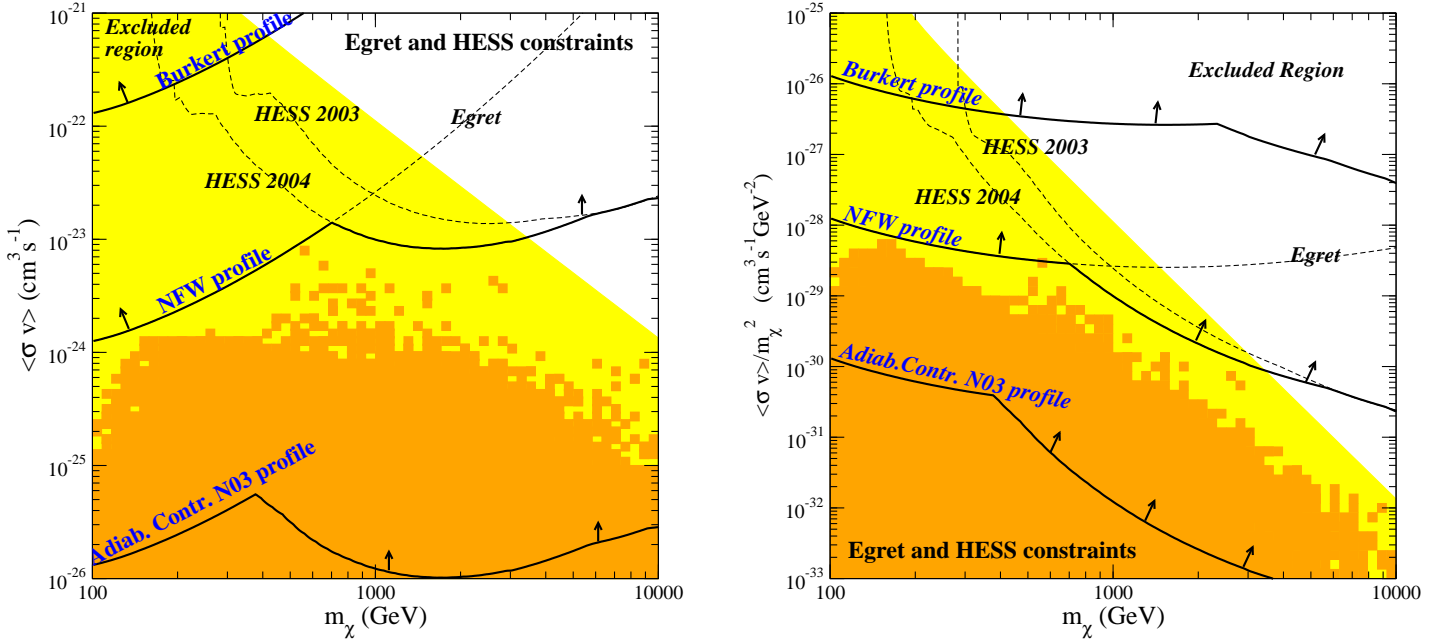

\begin{center}
\hspace*{-0.7cm}\mbox{\epsfig{file=mx_sv_egrethess.eps,height=8.5cm}\quad\quad\epsfig{file=mx_svm2_egrethess.eps,height=8.5cm}}
\end{center}
\caption{\it\small  The regions, in the $(m_\chi,\langle\sigma v\rangle)$ plane of supersymmetric models with non-rescaled cross section (yellow shaded area) and of models with a WMAP relic abundance or with a rescaled cross section (orange shaded area) excluded by the Egret data \cite{egret} and/or by the 2004 HESS data \cite{icrc_ripken} on the gamma ray flux from the center of the Galaxy, for three different halo models. The thin dashed lines indicate the shape of the Egret and HESS bounds, for the NFW profile, where one constraint is weaker than the other. We also indicate, again with a thin dashed line and for the NFW profile, the bound resulting from the 2003 HESS data. (Right): same as for the left panel, but for the ``supersymmetric factor''  $\langle\sigma v\rangle/m^2_\chi$.
}
\label{fig:maxsvegrethess}
\end{figure}
The results presented in this Section might be of relevance in the assessment of the scientific purposes of experimental devices sensitive to DM annihilation products, or in establishing whether the results of a given experiment provide, or not, any constraint on viable supersymmetric models. Further, our results can be of some use in order to construct {\em best case} scenarios for DM indirect detection signals, {\em e.g.} in evaluating the theoretically maximal supersymmetric DM induced gamma ray or neutrino fluxes from a given astrophysical system.

As an illustrative worked-out case study, we consider the halo-model dependent bounds put by the Egret \cite{egret} and by the HESS \cite{Aharonian:2004wa,icrc_rolland} data on the gamma ray flux from the galactic center on the planes in Fig.~\ref{fig:maxsv}. Given a DM density profile, one can conservatively require that the neutralino annihilations-induced gamma ray flux from the Galactic center alone does not exceed the fluxes measured by the Egret and by the HESS experiment, in any of the energy bins of the respective data sets. We performed this exercise in Fig.~\ref{fig:maxsvegrethess}, where we indicate with a solid black line the contours above which the parameter space is excluded at 95\% C.L. for three halo models: the Burkert profile \cite{burkert}, the NFW profile \cite{nfw} and the ACN03 profile \cite{n03,pierohalos}. We also indicate, with a thinner dashed line, the shape of the regions exclued by the 2004 HESS data (resp. by the Egret data), when Egret (resp. HESS) data place a stronger constraint. Furthermore, we indicate for reference, again with a dashed black line, the exclusion limits which could have been drawn out of the 2003 HESS data only: this gives an idea of the improvement on the constraints on DM physics gained with the 2004 HESS data.  All the exclusion limits have been computed in a model independent way, picking the most conservative option between a gauge boson dominated final state and a heavy quark final state.

Obviously, the HESS data only constrain neutralino masses larger than the minimal energy probed by the experiment, respectively $E_\gamma\approx 150$ GeV for the 2004 data and $E_\gamma\simeq 280$ GeV for the 2003 data. The shape of the regions excluded by Egret and by HESS highlights a nice {\em complementarity between satellite-borne experiments and ACTs} in the exploration of the viable supersymmetric parameter space. This complementarity will be further strengthened by the forthcoming space-borne GLAST and AMS experiments \cite{glast,ams} and by the new generation ACTs \cite{newact}. We also remark that, on fully general grounds, current gamma rays data only place constraints on models providing a WMAP relic abundance (or on models with a {\em rescaled} DM density) if the DM density profile is {\em cuspier than a NFW profile}. Finally, since it is hard to imagine the DM density profile in the Galaxy to be shallower than the cored Burkert profile (see also Fig.~\ref{fig:sigmaj}), the left panel shows that WIMP annihilation cross sections larger than $10^{-21}\ {\rm cm}^3{\rm s}^{-1}$ are in any case in conflict with the Egret data.

\section{Conclusions}\label{sec:conclusions}

For the ease of the reader, we collect below the main results of our analysis.

\begin{itemize}
\item Adopting a fully numerical approach, we determined the range of WIMP {\em masses} and pair annihilation {\em cross sections} needed to fit the ACT data on gamma rays from the galactic center in terms of DM annihilations. The 2004 HESS data are not compatible with the hypothesis of a WIMP annihilating into a pure gauge bosons final state as well as that of a Kaluza-Klein DM candidate; the option of a mixed $\tau^+\tau^-$-$b\bar b$ final state, viable within supersymmetric models, including some discussed in this paper (Sec.~\ref{sec:nuhm}), remains open, as a purely DM-annihilation interpretation of the HESS data. In the conservative ``{\em best spectral functions}'' approach, the WIMP annihilations interpretation of the HESS data is still valid, at 90\% C.L., in the mass range $6\lesssim m_\chi/{\rm TeV}\lesssim 30$ (Sec.~\ref{sec:act})
\item We determined the maximal neutralino mass compatible with a thermal neutralino relic abundance within the WMAP range in the context of two benchmark minimal models, mSUGRA and mAMSB. The largest possible masses we find are not compatible with the mass range of the HESS data, while an interpretation of the Cangaroo-II data in the context of the mAMSB model would require an extremely cuspy profile and an enhancement mechanism of the resulting thermal neutralino relic abundance (Sec.~\ref{sec:minimalmodels})
\item Enlarging the parameter space of mSUGRA and of mAMSB to general soft supersymmetry breaking Higgs mass terms, we showed that multi-TeV neutralinos are cosmologically allowed, and predicted to produce a detectable signal at future ton-sized direct detection experiments. Models with non-universal Higgs masses provide large LSP masses and pair annihilation cross sections, compatible with the ranges needed to fit the ACT data, and suitable gamma-rays spectral functions (Sec.~\ref{sec:nuhm})
\item We gave an analytical and a numerical estimate of the {\em largest possible neutralino pair annihilation cross section} in the general minimal supersymmetric extension of the standard model, both for models with and without gaugino mass unification. The largest cross sections in the MSSM occur for resonantly annihilating neutralinos with maximal gaugino-higgsino mixing (Sec.~\ref{sec:ann})
\item The inclusion of non-perturbative electro-weak resonant effects can produce pair annihilation cross sections larger than the maximal bounds we derived, but only for model-dependent narrow ranges of the neutralino mass. The resulting gamma rays spectral function, moreover, is not compatible with a DM annihilation origin for the HESS data (Sec.~\ref{sec:npew})
\item We then addressed the issue of the {\em largest possible neutralino mass} in the MSSM, compatible with a ``WMAP'' thermal relic abundance. We studied the consequences of a full non-perturbative treatment of QCD effects in the pair annihilation cross section of a viable neutralino coannihilating partner, the gluino. We showed that in this context, considering extreme non-perturbative scenarios, a neutralino with a mass much larger than 100 TeV, hence well beyond the $s$-wave unitarity limit, can yield a low enough thermal relic abundance (Sec.~\ref{sec:gluino})
\item We numerically assessed the maximal neutralino pair annihilation cross section and the largest possible ``supersymmetric factor'' ($\langle\sigma v\rangle/m^2_\chi$), relevant for any indirect DM detection rate, for models with a WMAP relic abundance, or with a ``rescaled'' cross section (see Eq.~(\ref{eq:rescale})). The results we presented provide a conservative {\em best-case scenario} for indirect supersymmetric DM detection. As a particular instance, we showed how available data on the flux of gamma rays from the galactic center, as measured by Egret and HESS, constrain the viable supersymmetric parameter space (Sec.~\ref{sec:outlook})
\end{itemize}


\vspace*{1cm}
\noindent{ {\bf Acknowledgments} } \\
\noindent
I would like to thank Howard Baer and Piero Ullio for useful suggestions and comments. This work was supported in part by the U.S. Department of Energy under contract number DE-FG02-97ER41022.



\begin{thebibliography}{99}
\small


\bibitem{Spergel:2003cb}
D.~N.~Spergel {\it et al.}  [WMAP Collaboration],
Astrophys.\ J.\ Suppl.\  {\bf 148} (2003) 175
[arXiv:astro-ph/0302209].

\bibitem{Bertone:2004pz}
  G.~Bertone, D.~Hooper and J.~Silk,
  Phys.\ Rept.\  {\bf 405} (2005) 279
  [arXiv:hep-ph/0404175].

\bibitem{Bergstrom:2000pn}
  L.~Bergstrom,
  Rept.\ Prog.\ Phys.\  {\bf 63} (2000) 793
  [arXiv:hep-ph/0002126].

\bibitem{Baltz:2004tj}
  E.~A.~Baltz,
  eConf {\bf C040802}, L002 (2004)
  [arXiv:astro-ph/0412170].


\bibitem{Munoz:2003gx}
  C.~Munoz,
  Int.\ J.\ Mod.\ Phys.\ A {\bf 19} (2004) 3093
  [arXiv:hep-ph/0309346].



\bibitem{ref:neutrinos}
J.~Silk, K.~Olive and M.~Srednicki,
Phys.\ Rev.\ Lett.\ {\bf 55}, 257 (1985);
K.~Freese,
Phys.\ Lett.\ B {\bf 167}, 295 (1986);
L.~M.~Krauss, M.~Srednicki and F.~Wilczek,
Phys.\ Rev.\ D {\bf 33}, 2079 (1986);
T.~K.~Gaisser, G.~Steigman and S.~Tilav,
Phys.\ Rev.\ D {\bf 34}, 2206 (1986);
L.~Bergstrom, J.~Edsjo and P.~Gondolo,
Phys.\ Rev.\ D {\bf 58}, 103519 (1998);
D.~Hooper and J.~Silk,
New J.\ Phys.\  {\bf 6}, 023 (2004)
[arXiv:hep-ph/0311367].

\bibitem{ref:gammarays}
S.~Rudaz and F.~W.~Stecker,
Astrophys.\ J.\  {\bf 325}, 16 (1988);
H.~U.~Bengtsson, P.~Salati and J.~Silk,
Nucl.\ Phys.\ B {\bf 346}, 129 (1990);
V.~Berezinsky, A.~Bottino and G.~Mignola,
Phys.\ Lett.\ B {\bf 325}, 136 (1994)
[arXiv:hep-ph/9402215];
F.~Stoehr, S.~D.~M.~White, V.~Springel, G.~Tormen and N.~Yoshida,
Mon.\ Not.\ Roy.\ Astron.\ Soc.\  {\bf 345}, 1313 (2003)
[arXiv:astro-ph/0307026];


\bibitem{Bergstrom:1997fj}
L.~Bergstrom, P.~Ullio and J.~H.~Buckley,
Astropart.\ Phys.\  {\bf 9}, 137 (1998)
[arXiv:astro-ph/9712318].


\bibitem{Bergstrom:2001jj}
L.~Bergstrom, J.~Edsjo and P.~Ullio,
Phys.\ Rev.\ Lett.\  {\bf 87}, 251301 (2001)
[arXiv:astro-ph/0105048].


\bibitem{ref:positrons}
M.~Kamionkowski and M.~S.~Turner,
Phys.\ Rev.\ D {\bf 43}, 1774 (1991);
E.~A.~Baltz and J.~Edsjo,
Phys.\ Rev.\ D {\bf 59} (1999) 023511
[arXiv:astro-ph/9808243];
G.~L.~Kane, L.~T.~Wang and J.~D.~Wells,
Phys.\ Rev.\ D {\bf 65}, 057701 (2002);
D.~Hooper, J.~E.~Taylor and J.~Silk,
Phys.\ Rev.\ D, in press [arXiv:hep-ph/0312076].

\bibitem{ref:antiprotons}
J.~Silk and M.~Srednicki,
Phys.\ Rev.\ Lett.\  {\bf 53}, 624 (1984);
F.~W.~Stecker, S.~Rudaz and T.~F.~Walsh,
Phys.\ Rev.\ Lett.\  {\bf 55}, 2622 (1985);
F.~Donato, N.~Fornengo, D.~Maurin, P.~Salati and R.~Taillet,
Phys.\ Rev.\ D {\bf 69}, 063501 (2004)
[arXiv:astro-ph/0306207].

\bibitem{ref:dbar}
  F.~Donato,
  Nucl.\ Phys.\ Proc.\ Suppl.\  {\bf 87} (2000) 445.
  K.~Mori, C.~J.~Hailey, E.~A.~Baltz, W.~W.~Craig, M.~Kamionkowski, W.~T.~Serber and P.~Ullio,
  Astrophys.\ J.\  {\bf 566} (2002) 604
  [arXiv:astro-ph/0109463].
  S.~Profumo and P.~Ullio,
  JCAP {\bf 0407} (2004) 006
  [arXiv:hep-ph/0406018].

\bibitem{Colafrancesco:2005ji}
  S.~Colafrancesco, S.~Profumo and P.~Ullio,
  arXiv:astro-ph/0507575.


\bibitem{Jungman:1995df}
  G.~Jungman, M.~Kamionkowski and K.~Griest,
  Phys.\ Rept.\  {\bf 267} (1996) 195
  [arXiv:hep-ph/9506380].


\bibitem{Chung:2003fi}
  D.~J.~H.~Chung, L.~L.~Everett, G.~L.~Kane, S.~F.~King, J.~Lykken and L.~T.~Wang,
  Phys.\ Rept.\  {\bf 407} (2005) 1
  [arXiv:hep-ph/0312378].


\bibitem{Hisano:2004ds}
  J.~Hisano, S.~Matsumoto, M.~M.~Nojiri and O.~Saito,
  Phys.\ Rev.\ D {\bf 71} (2005) 063528
  [arXiv:hep-ph/0412403].


\bibitem{Tsuchiya:2004wv}
  K.~Tsuchiya {\it et al.}  [CANGAROO-II Collaboration],
  Astrophys.\ J.\  {\bf 606} (2004) L115
  [arXiv:astro-ph/0403592].


\bibitem{Aharonian:2004wa}
  F.~Aharonian {\it et al.}  [The HESS Collaboration],
  arXiv:astro-ph/0408145.


\bibitem{hess2004}  L.~Rolland, talk given at the conference ``Very High Energy Phenomena in the Universe'' La Thuile, Italy (March 12-19, 2005), and F.~Aharonian and W.~Benbow, talks given at the conference ``TeV Particle Astrophysics '', 13-15 July 2005. Fermilab, Batavia, IL, USA.

\bibitem{icrc_rolland}
L.~Rolland and J.~Hinton for the HESS collaboration, Proceedings of the 29th International Cosmic Ray Conference, Pune, India (2005) ({\tt http://icrc2005.tifr.res.in}).

\bibitem{icrc_ripken}
J.~Ripken, D.~Horns, L.~Rolland and J.~Hinton for the HESS collaboration, Proceedings of the 29th International Cosmic Ray Conference, Pune, India (2005) ({\tt http://icrc2005.tifr.res.in}).

\bibitem{canghessdisc} M. Mori in Procs. of 28th
	ICRC (2003), Vol. 5, ed. T. Kajita, Y. Asaka, A. Kawachi, Y. Matsubara, and M. Sasaki 
(Tokyo: Universal Academy Press); N. Komin et al. (2004) in F.A. Aharonian \& H.J. V\"olk (eds.) AIP Conference Series in press; D. Berge et al. (2004) in F.A. Aharonian \& H.J. V\"olk (eds.) AIP Conference Series in press.



\bibitem{Aharonian:2004jr}
  F.~Aharonian and A.~Neronov,
  Astrophys.\ J.\  {\bf 619}, 306 (2005)
  [arXiv:astro-ph/0408303].

\bibitem{maeda}
 Y.~Maeda {\em et al.}, ApJ, {\bf 570}, 671 (2002).

\bibitem{Fatuzzo:2003nw}
  M.~Fatuzzo and F.~Melia,
  Astrophys.\ J.\  {\bf 596} (2003) 1035
  [arXiv:astro-ph/0302607].

\bibitem{Hooper:2004vp}
  D.~Hooper, I.~de la Calle Perez, J.~Silk, F.~Ferrer and S.~Sarkar,
  JCAP {\bf 0409}, 002 (2004)
  [arXiv:astro-ph/0404205].


\bibitem{Kosack:2004ri}
  K.~Kosack {\it et al.}  [The VERITAS Collaboration],
  Astrophys.\ J.\  {\bf 608}, L97 (2004)
  [arXiv:astro-ph/0403422].


\bibitem{Horns:2004bk}
  D.~Horns,
  Phys.\ Lett.\ B {\bf 607}, 225 (2005)
  [Erratum-ibid.\ B {\bf 611}, 297 (2005)]
  [arXiv:astro-ph/0408192].


\bibitem{Hooper:2004fh}
  D.~Hooper and J.~March-Russell,
  Phys.\ Lett.\ B {\bf 608}, 17 (2005)
  [arXiv:hep-ph/0412048].


\bibitem{Bergstrom:2004cy}
  L.~Bergstrom, T.~Bringmann, M.~Eriksson and M.~Gustafsson,
  Phys.\ Rev.\ Lett.\  {\bf 94} (2005) 131301
  [arXiv:astro-ph/0410359];  L.~Bergstrom, T.~Bringmann, M.~Eriksson and M.~Gustafsson,
  JCAP {\bf 0504} (2005) 004
  [arXiv:hep-ph/0412001].

\bibitem{nfw}
  J.~F.~Navarro, C.~S.~Frenk and S.~D.~M.~White,
  Astrophys.\ J.\  {\bf 462} (1996) 563
  [arXiv:astro-ph/9508025];
  J.S.~Bullock et al., MNRAS {\bf 321} (2001) 559;
  V.R.~Eke, J.F.~Navarro and M.~Steinmetz, Astrophys. J. {\bf 554}
  (2001) 114.

\bibitem{egret}
  H.~A.~Mayer-Hasselwander {\it et al.}, Astron. \& Astrophys. {\bf 335} (1998) 161;
  S.~D.~Hunger {\it et al.},
  Astrophys.\ J.\  {\bf 481}, 205 (1997).

\bibitem{Mambrini:2005vk}
  Y.~Mambrini, C.~Munoz, E.~Nezri and F.~Prada,
  arXiv:hep-ph/0506204.

\bibitem{Fornengo:2004kj}
  N.~Fornengo, L.~Pieri and S.~Scopel,
  Phys.\ Rev.\ D {\bf 70}, 103529 (2004)
  [arXiv:hep-ph/0407342].

\bibitem{pythia}
T.~Sj\"{o}strand, fixxx  cpc{82}{1994}{74};
T.~Sj\"{o}strand, {\em PYTHIA 5.7 and JETSET 7.4. Physics and Manual},
CERN-TH.7112/93, arXiv:hep-ph/9508391 (revised version).

\bibitem{Gondolo:2004sc}
P.~Gondolo, J.~Edsjo, P.~Ullio, L.~Bergstrom, M.~Schelke and E.~A.~Baltz,
JCAP {\bf 0407} (2004) 008
[arXiv:astro-ph/0406204].

\bibitem{Servant:2002aq}
  G.~Servant and T.~M.~P.~Tait,
  Nucl.\ Phys.\ B {\bf 650} (2003) 391
  [arXiv:hep-ph/0206071].





\bibitem{Bergstrom:2005ss}
  L.~Bergstrom, T.~Bringmann, M.~Eriksson and M.~Gustafsson,
  arXiv:hep-ph/0507229.

\bibitem{burkert}
  A.~Burkert, Astrophys. J. {\bf 447} (1995) L25;
  P.~Salucci and A.~Burkert, Astrophys. J. {\bf 537} (2000) L9.

\bibitem{blumental}
  G.R.~Blumental, S.M. Faber, R.~Flores and J.R.~Primack,  
  Astrophys. J. {\bf 301} (1986) 27.

\bibitem{n03}
 J.F.~Navarro et al., MNRAS (2004) in press, astro-ph/0311231.

\bibitem{pierohalos} P.~Ullio, proceedings of the Third International Conference on Frontier Science, Monteporzio Catone (RM), Italy, ed. by Frascati Physics Series.

\bibitem{rev} For reviews of SUSY phenomenology, see, S. P.~Martin,
hep-ph/9709356; M.~Drees, hep-ph/9611409; X.~Tata, hep-ph/9706307;
S.~Dawson, hep-ph/9712464.
%


\bibitem{Baer:2000gf}
  H.~Baer, M.~A.~Diaz, P.~Quintana and X.~Tata,
  JHEP {\bf 0004} (2000) 016
  [arXiv:hep-ph/0002245];
  S.~Profumo,
  JHEP {\bf 0306}, 052 (2003)
  [arXiv:hep-ph/0306119].

\bibitem{Profumo:2004at}
  S.~Profumo and C.~E.~Yaguna,
  Phys.\ Rev.\ D {\bf 70}, 095004 (2004)
  [arXiv:hep-ph/0407036].

\bibitem{msugra} A.~Chamseddine, R.~Arnowitt and P.~Nath, 
Phys. Rev. Lett. {\bf 49} (1982) {970};
R.~Barbieri, S.~Ferrara and C.~Savoy, 
Phys. Lett. {\bf B119} (1982) {343};
L.~J.~Hall, J.~Lykken and S.~Weinberg, Phys.\ Rev.\ D,{\bf 27} (1983) {2359};
for reviews, see H.~P.~Nilles, Phys. Rept. {\bf 110} (1984) {1} and
  P.~Nath,
  arXiv:hep-ph/0307123.
%
\bibitem{Giudice:1998xp}
G.~F.~Giudice, M.~A.~Luty, H.~Murayama and R.~Rattazzi,
JHEP {\bf 9812} (1998) 027
[arXiv:hep-ph/9810442].

\bibitem{Randall:1998uk}
L.~Randall and R.~Sundrum,
Nucl.\ Phys.\ B {\bf 557} (1999) 79
[arXiv:hep-th/9810155].

\bibitem{Gherghetta:1999sw}
T.~Gherghetta, G.~F.~Giudice and J.~D.~Wells,
Nucl.\ Phys.\ B {\bf 559} (1999) 27
[arXiv:hep-ph/9904378].


\bibitem{Feng:1999hg}
J.~L.~Feng and T.~Moroi,
Phys.\ Rev.\ D {\bf 61} (2000) 095004
[arXiv:hep-ph/9907319].

\bibitem{bulk} See {\it e.g.}, H. Baer and M. Brhlik, \prd{53}{1996}{597}.
%
\bibitem{stau} J. Ellis, T. Falk and K. Olive, 
\plb{444}{1998}{367}; J. Ellis, T. Falk, K. Olive and M. Srednicki,
\app{13}{2000}{181}; R.~Arnowitt, B.~Dutta and Y.~Santoso,
\npb{606}{2001}{59}.
%
\bibitem{Afunnel} M. Drees and M. Nojiri, \prd{47}{1993}{376}; 
H. Baer and M. Brhlik, \prd{57}{1998}{567} and Ref. \cite{bulk};
H. Baer, M. Brhlik, M. Diaz, J. Ferrandis, P. Mercadante,
P. Quintana and X. Tata, \prd{63}{2001}{015007};
J. Ellis, T. Falk, G. Ganis, K. Olive and M. Srednicki,
\plb{510}{2001}{236}; L. Roszkowski, R. Ruiz de Austri and T. Nihei,
\jhep{0108}{024}{2001}. 
A. Lahanas and V. Spanos, \epjc{23}{2002}{185}.
%
\bibitem{hb_fp} K. L. Chan, U. Chattopadhyay and P. Nath, 
\prd{58}{1998}{096004}.
J.~Feng, K.~Matchev and T.~Moroi, \prl{84}{2000}{2322} and 
\prd{61}{2000}{075005}; see also 
H. Baer, C. H. Chen, F. Paige and X. Tata, \prd{52}{1995}{2746} and 
\prd{53}{1996}{6241}; H. Baer, C. H. Chen, M. Drees, F. Paige and X. Tata, 
\prd{59}{1999}{055014}.
%

\bibitem{Baer:2005ky}
  H.~Baer, T.~Krupovnickas, S.~Profumo and P.~Ullio,
  arXiv:hep-ph/0507282.
%
\bibitem{Boehm:1999bj}
  C.~Boehm, A.~Djouadi and M.~Drees,
  Phys.\ Rev.\ D {\bf 62}, 035012 (2000)
  [arXiv:hep-ph/9911496].

\bibitem{stop}
J.~R.~Ellis, K.~A.~Olive and Y.~Santoso, \app{18}{2003}{395};


\bibitem{Edsjo:2003us}
  J.~Edsjo, M.~Schelke, P.~Ullio and P.~Gondolo,
  JCAP {\bf 0304}, 001 (2003)
  [arXiv:hep-ph/0301106].

\bibitem{drees_h} 
R. Arnowitt and P. Nath, \prl{70}{1993}{3696};
H. Baer and M. Brhlik, Ref. \cite{bulk};  
A. Djouadi, M. Drees and J. Kneur, hep-ph/0504090.
%

\bibitem{Profumo:2004qt}
  S.~Profumo, K.~Sigurdson, P.~Ullio and M.~Kamionkowski,
  Phys.\ Rev.\ D {\bf 71}, 023518 (2005)
  [arXiv:astro-ph/0410714].

\bibitem{splitsusy}
  G.~F.~Giudice and A.~Romanino,
  Nucl.\ Phys.\ B {\bf 699}, 65 (2004)
  [Erratum-ibid.\ B {\bf 706}, 65 (2005)]
  [arXiv:hep-ph/0406088];
  A.~Masiero, S.~Profumo and P.~Ullio,
  Nucl.\ Phys.\ B {\bf 712}, 86 (2005)
  [arXiv:hep-ph/0412058].

\bibitem{quint}
%
P.~Salati,
[arXiv:astro-ph/0207396];
F.~Rosati,
Phys.\ Lett.\ B {\bf 570} (2003) 5
[arXiv:hep-ph/0302159];
%
S.~Profumo and P.~Ullio,
JCAP {\bf 0311}, 006 (2003)
[arXiv:hep-ph/0309220];
%
\bibitem{Kamionkowski:1990ni}
M.~Kamionkowski and M.~S.~Turner,
Phys.\ Rev.\ D {\bf 42} (1990) 3310;

\bibitem{bdj}
R.~Catena, N.~Fornengo, A.~Masiero, M.~Pietroni and F.~Rosati,
arXiv:astro-ph/0403614;
%
\bibitem{Profumo:2004ex}
S.~Profumo and P.~Ullio,
Proceedings of the {\em 39th Rencontres de Moriond Workshop on
Exploring the Universe: Contents and Structures of the Universe},
La Thuile, Italy, 28 Mar - 4 Apr 2004, ed. by J. Tran Thanh Van [arXiv:hep-ph/0305040].
%

\bibitem{nonth}
B.~Murakami and J.~D.~Wells,
Phys.\ Rev.\ D {\bf 64}, 015001 (2001);
T.~Moroi and L.~Randall,
Nucl.\ Phys.\ B {\bf 570}, 455 (2000);
M.~Fujii and K.~Hamaguchi,
Phys.\ Lett.\ B {\bf 525}, 143 (2002);
M.~Fujii and K.~Hamaguchi,
Phys.\ Rev.\ D {\bf 66}, 083501 (2002);
R.~Jeannerot, X.~Zhang and R.~H.~Brandenberger,
JHEP {\bf 9912}, 003 (1999);
W.~B.~Lin, D.~H.~Huang, X.~Zhang and R.~H.~Brandenberger,
Phys.\ Rev.\ Lett.\  {\bf 86}, 954 (2001);
\bibitem{nonuniversalscalar}
A.~Datta, M.~Guchait, N.~Parua, {\em Phys. Lett.} {\bf B 395} (1997) 54;
A.~Datta, A.~Datta, M.K.~Parida, {\em Phys. Lett.} {\bf B 431} (1998) 347;
E.~Accomando, R.~Arnowitt, B.~Dutta and Y.~Santoso, {\em Nucl. Phys.} {\bf B585} (2000) 124;
S.~Codoban, M.~Jurcisin and D.~Kazakov, {\em Phys. Lett.} {\bf B 477} (2000) 223.
H.~Baer, C.~Balazs, S.~Hesselbach, J.K.~Mizukoshi and X.~Tata, {\em Phys. Rev.} {\bf D 63} (2001) 095008;
J.~Ellis, T.~Falk, K.A.~Olive and Y.~Santoso, \emph{Nucl. Phys.} \textbf{B652} (2003) 259;
  S.~Profumo,
  Phys.\ Rev.\ D {\bf 68} (2003) 015006
  [arXiv:hep-ph/0304071] and
  arXiv:hep-ph/0305040.

\bibitem{nugm}
A.~Birkedal-Hansen, B.D.~Nelson, {\tt hep-ph/0211071}; S.I.~Bityukov, N.V.~Krasnikov, {\em Phys. Atom. Nucl.} {\bf 65} (2002) 1341;
H.~Baer, C.~Balazs, A.~Belyaev, R.~Dermisek, A.~Mafi and A.~Mustafayev, {\em JHEP} {\bf 0205} (2002) 061;
N.~Chamoun, C.S.~Huang, C.~Liu and X.H.~Wu, {\em Nucl. Phys.} {\bf B624} (2002) 81;  C.~Balazs and R.~Dermisek,
  JHEP {\bf 0306}, 024 (2003)
  [arXiv:hep-ph/0303161];
  H.~Baer, A.~Mustafayev, E.~K.~Park and S.~Profumo,
  JHEP {\bf 0507} (2005) 046
  [arXiv:hep-ph/0505227] and 
  H.~Baer, T.~Krupovnickas, A.~Mustafayev, E.~K.~Park, S.~Profumo and X.~Tata, in preparation.


\bibitem{nuhm}
%
V.~Berezinsky, A. Bottino, J. Ellis, N. Fornengo,
G. Mignola and S. Scopel, \app{5}{1996}{1};
%
V.~Berezinsky, A. Bottino, J. Ellis, N. Fornengo,
G. Mignola and S. Scopel, \app{5}{1996}{333}; 
N. Fornengo, \npps{52A}{1997}{239};
%
R. Arnowitt and P. Nath, \prd{56}{1997}{2820};
%
A. Bottino, F. Donato, N. Fornengo and S. Scopel, 
\prd{59}{1999}{095004};
%
A. Bottino, F. Donato, N. Fornengo and S. Scopel, 
\prd{63}{2001}{125003};
%
J. Ellis, K. Olive and Y. Santoso, \plb{539}{2002}{107};
J. Ellis, T. Falk, K. Olive and Y. Santoso, \npb{652}{2003}{259};
%
J. Ellis, A. Ferstl, K. Olive and Y. Santoso, 
\prd{67}{2003}{123502};
%
V. Barger, F. Halzen, D. Hooper and C. Kao,
\prd{65}{2002}{075022};
%
D.G. Cerdeno, E. Gabrielli, M.E. Gomez and C. Munoz,
\jhep{0306}{2003}{030}; D.G. Cerdeno, C. Munoz, \jhep{0410}{2004}{015};
%
Y. Mambrini and C. Munoz, 
\jcap{0410}{2004}{003} and hep-ph/0407158
%
  H.~Baer, A.~Mustafayev, S.~Profumo, A.~Belyaev and X.~Tata,
  Phys.\ Rev.\ D {\bf 71} (2005) 095008
  [arXiv:hep-ph/0412059].

\bibitem{Baer:2005bu}
  H.~Baer, A.~Mustafayev, S.~Profumo, A.~Belyaev and X.~Tata,
  JHEP {\bf 0507} (2005) 065
  [arXiv:hep-ph/0504001].

\bibitem{Aprile:2004ey}
E.~Aprile {\it et al.},
arXiv:astro-ph/0407575.

\bibitem{Gondolo:1990dk}
  P.~Gondolo and G.~Gelmini,
  Nucl.\ Phys.\ B {\bf 360}, 145 (1991).


\bibitem{Edsjo:1997bg}
  J.~Edsjo and P.~Gondolo,
  Phys.\ Rev.\ D {\bf 56} (1997) 1879
  [arXiv:hep-ph/9704361].


\bibitem{Eidelman:2004wy}
  S.~Eidelman {\it et al.}  [Particle Data Group],
  Phys.\ Lett.\ B {\bf 592} (2004) 1.


\bibitem{Cheng:1998hc}
  H.~C.~Cheng, B.~A.~Dobrescu and K.~T.~Matchev,
  Nucl.\ Phys.\ B {\bf 543} (1999) 47
  [arXiv:hep-ph/9811316].

\bibitem{Hisano:2002fk}
  J.~Hisano, S.~Matsumoto and M.~M.~Nojiri,
  Phys.\ Rev.\ D {\bf 67}, 075014 (2003)
  [arXiv:hep-ph/0212022].


\bibitem{Boudjema:2005hb}
  F.~Boudjema, A.~Semenov and D.~Temes,
  arXiv:hep-ph/0507127.


\bibitem{Giudice:1995qk}
  G.~F.~Giudice and A.~Pomarol,
  Phys.\ Lett.\ B {\bf 372}, 253 (1996)
  [arXiv:hep-ph/9512337].

\bibitem{GriestSeckel}
K.~Griest and D.~Seckel, \emph{Phys. Rev.} \textbf{D 43} (1991) 3191.


\bibitem{Griest:1989wd}
  K.~Griest and M.~Kamionkowski,
  Phys.\ Rev.\ Lett.\  {\bf 64} (1990) 615.


\bibitem{Profumo:2004wk}
  S.~Profumo and C.~E.~Yaguna,
  Phys.\ Rev.\ D {\bf 69}, 115009 (2004)
  [arXiv:hep-ph/0402208].

\bibitem{micromegas}
G. Belanger, F. Boudjema, A. Pukhov and A. Semenov,  hep-ph/0405253; G. Belanger, F. Boudjema, A. Pukhov and A. Semenov,  Comput. Phys. Commun. 149 (2002) 103; hep-ph/0112278.



\bibitem{Baer:1998pg}
  H.~Baer, K.~m.~Cheung and J.~F.~Gunion,
  Phys.\ Rev.\ D {\bf 59}, 075002 (1999)
  [arXiv:hep-ph/9806361].


\bibitem{oII}
V.S. Kaplunovsky and J. Louis, Phys.\ Lett. {\bf B306} 269 (1993) ; 
A. Brignole, L.E. Ibanez and C. Munoz, Nucl. Phys. B {\bf 422} 125 (1994) ,
[E: {\bf B436} (1995) 747]; CERN-TH/97-143 [hep-ph/9707209];
C.H. Chen, M. Drees, and J.F. Gunion,  Phys.\ Rev. D {\bf 55} 330 (1997).

\bibitem{Arkani-Hamed:2004fb}
  N.~Arkani-Hamed and S.~Dimopoulos,
  JHEP {\bf 0506}, 073 (2005)
  [arXiv:hep-th/0405159];
  N.~Arkani-Hamed, S.~Dimopoulos, G.~F.~Giudice and A.~Romanino,
  Nucl.\ Phys.\ B {\bf 709}, 3 (2005)
  [arXiv:hep-ph/0409232].

\bibitem{othergluino}
S. Raby, \prd{56}{1997}{2852}; Phys.\ Lett.\ B {\bf 422} (1998) 158;
C.B. Dover, T.K. Gaisser and G. Steigman, \prl{42}{1979}{1117};
S. Wolfram, \plb{82}{1979}{65};
R.N. Mohapatra and S. Nussinov, \prd{57}{1997}{1940}.


\bibitem{Jedamzik:2004ip}
  K.~Jedamzik,
  Phys.\ Rev.\ D {\bf 70}, 083510 (2004)
  [arXiv:astro-ph/0405583] and
  K.~Jedamzik,
  Phys.\ Rev.\ D {\bf 70}, 063524 (2004)
  [arXiv:astro-ph/0402344].

\bibitem{Bottino:2005xy}
  A.~Bottino, F.~Donato, N.~Fornengo and P.~Salati,
  arXiv:hep-ph/0507086.


\bibitem{glast}
J. E. McEnery, I. V. Moskalenko and J. F. Ormes,
  astro-ph/0406250.

\bibitem{ams}
The {\tt AMS} Collaboration, S.~P.~Ahlen {\it et al.}, Nucl.\
Instrum.\ Meth.\
  A {\bf 350}, 351 (1994); J.~Alcaraz {\it et al.}, Nucl.\ Instrum.\ Meth.\ A
  {\bf 478}, 119 (2002);  A.~Jacholkowska {\it et al.},
  arXiv:astro-ph/0508349.

\bibitem{newact}
K.~Tsuchiya {\it et al.}, The CANGAROO Collaboration,
Astrophys. J. Lett. {\bf
  606}; astro-ph/0403592; J.A. Hinton, The HESS Collaboration, New Astron.Rev. {\bf
48} (2004) 331; C.~Baixeras, Nucl.Phys.B (Proc.Suppl.) {\bf 114} (2003) 247; T.C.~Weekes {\it et al.}, Astropart. Phys. {\bf 17} (2002) 221. 

\end{thebibliography}
\end{document}